\documentclass[aps,reprint,nofootinbib,nobibnotes,notitlepage,superscriptaddress,onecolumn,prd,
 amsmath,amssymb
]{revtex4-2}

\usepackage[caption=false]{subfig}
\usepackage{braket}
\usepackage{float}
\usepackage{lipsum}
\usepackage{graphicx}% include figure files
\usepackage{dcolumn}% align table columns on decimal point
\usepackage{bm}% bold math
\usepackage{natbib}
\usepackage{hyperref}
\hypersetup{
	colorlinks = true,
    linkcolor = Red,
    urlcolor  = Red,
    citecolor = Red
}

\usepackage{microtype}

\usepackage{verbatim}
\usepackage[amssymb]{SIunits}
\usepackage{tabularx}
\usepackage[dvipsnames]{xcolor}
\usepackage{wasysym}

\usepackage{tikz}

\def\be{\begin{equation}}
\def\ee{\end{equation}}

\usepackage{color}
\definecolor{darkgreen}{RGB}{0,120,0}

\definecolor{darkgreen}{RGB}{0,120,0}
\newcommand{\resub}[1]{{#1}}%\color{darkgreen}{#1}}}

\newcommand{\delD}[1]{(2\pi)^3\delta_\mathrm{D}\left({#1}\right)}

\newcommand{\av}[1]{\left\langle{#1}\right\rangle} 

\newcommand{\vk}{\vec k}
\newcommand{\hk}{\hat{\vec k}}

\newcommand{\vx}{\vec x}

\newcommand{\Si}{\mathsf{S}^{-1}}

\newcommand{\hn}{\hat{\vec n}}

\newcommand{\tj}[6]{\begin{pmatrix} {#1} & {#2} & {#3}\\ {#4} & {#5} & {#6}\end{pmatrix}}

\def\beq{\begin{eqnarray}}
\def\eeq{\end{eqnarray}}
\let\vec\mathbf

\usepackage{empheq}

\def\mnras{\rm{MNRAS}}

\def\aap{\rm{A\&A}}

\begin{document}

\title{Testing Graviton Parity and Gaussianity with \textit{Planck} $T$-, $E$- and $B$-mode Bispectra}
%\title{Constraining Tensor Bispectra with \textit{Planck} $T$, $E$ and $B$ modes}
%\date{\today}

\author{Oliver~H.\,E.~Philcox}
\email{ohep2@cantab.ac.uk}
\affiliation{Department of Physics,
Columbia University, New York, NY 10027, USA}
\affiliation{Simons Society of Fellows, Simons Foundation, New York, NY 10010, USA}

\author{Maresuke Shiraishi}
\email{shiraishi\_maresuke@rs.sus.ac.jp}
\affiliation{School of General and Management Studies, Suwa University of Science, Chino, Nagano
391-0292, Japan}

\begin{abstract} 
    \noindent Many inflationary theories predict a non-Gaussian spectrum of primordial tensor perturbations, sourced from non-standard vacuum fluctuations, modified general relativity or new particles such as gauge fields. Several such models also predict a chiral spectrum in which one polarization state dominates. In this work, we place constraints on the non-Gaussianity and parity properties of primordial gravitational waves utilizing the \textit{Planck} PR4 temperature and polarization dataset. Using recently developed quasi-optimal bispectrum estimators, we compute binned parity-even and parity-odd bispectra for all combinations of CMB $T$-, $E$- and $B$-modes with $2\leq \ell<500$, and perform both blind tests, sensitive to arbitrary three-point functions, and targeted analyses of a well-motivated equilateral gravitational wave template (sourced by gauge fields), with amplitude $f_{\rm NL}^{ttt}$. This is the first time $B$-modes have been included in primordial non-Gaussianity analyses; they are found to strengthen constraints on the parity-even sector by $\simeq 30\%$ and dominate the parity-odd bounds, without inducing bias. We report no detection of non-Gaussianity (of either parity), with the template amplitude constrained to $f_{\rm NL}^{ttt}=900\pm 700$ (stable with respect to a number of analysis variations), compared to $1300\pm1200$ in \textit{Planck} 2018. The methods applied herein can be reapplied to upcoming CMB datasets such as LiteBIRD, with the inclusion of $B$-modes poised to dramatically improve future bounds on tensor non-Gaussianity.
\end{abstract}

\maketitle

\section{Introduction}\label{sec: intro}

\vskip 4pt 

\noindent High-precision measurements of the cosmic microwave background (CMB) have provided fruitful information on primordial scalar fluctuations, supporting both the $\Lambda$CDM model and an inflationary epoch in the very early Universe. In contrast, experiments have provided little insight into the primordial tensor sector, due to the non-detection of early Universe gravitational waves (GWs), despite numerous searches \citep{2020A&A...641A...6P,Planck:2018jri} (see \citep{Kamionkowski:2015yta} for a review).

Primordial GWs have been widely regarded as a smoking gun of cosmic inflation since they are naturally generated by quantum fluctuations of the inflationary vacuum. However, they may also be sourced by other mechanisms. For instance, if gauge fields exist and do not decay during inflation, GWs can be sourced by gravitational interactions \citep[e.g.,][]{Watanabe:2010fh,Barnaby:2012xt}. Their production becomes further efficient if the gauge fields couple to axions \citep{Lue:1998mq,Sorbo:2011rz,Barnaby:2011vw,Barnaby:2012xt,Cook:2013xea,Namba:2015gja,Dimastrogiovanni:2016fuu} (see also \citep{Maleknejad:2012fw,Komatsu:2022nvu} for reviews). Furthermore, gravitational waves could be a signature of, e.g., non-standard primordial vacuum states (such as $\alpha$-vacua) \citep{Gong:2023kpe,Kanno:2022mkx}, non-attractor phases in inflation \citep{Ozsoy:2019slf}, additional scalar-tensor and derivative couplings \citep{Gao:2011vs,Gao:2012ib,DeLuca:2019jzc,Maldacena:2011nz}, or Chern-Simons modified gravity \citep{Lue:1998mq,
Alexander:2004wk,Soda:2011am,Shiraishi:2011st,Bartolo:2017szm} (see \citep{Guzzetti:2016mkm} and references therein for other scenarios). Many such contributions can be efficiently modeled in a UV-physics-agnostic manner with Effective Field Theory of Inflation, often invoking bootstrap methods \citep{Cabass:2021fnw,Cabass:2022jda,Bordin:2020eui,Cabass:2021iii,Pajer:2020wxk}. Being able to distinguish between these contributions and those intrinsic to a simple Bunch-Davies vacuum is thus an important issue.

Under standard inflationary assumptions, GWs sourced by the vaccum have a statistically monotonic spectrum, which is both parity-symmetric and (approximately) Gaussian \citep{Maldacena:2002vr}. More complex models involving, for example, axion-gauge field interactions can source non-Gaussian and/or parity-violating distributions, whose particular form encodes the microphysics of the early Universe \citep{Lue:1998mq,Sorbo:2011rz,Barnaby:2011vw,Barnaby:2012xt,Cook:2013xea,Namba:2015gja,Dimastrogiovanni:2016fuu,Agrawal:2017awz,Ozsoy:2019slf,Gong:2023kpe,Shiraishi:2016yun,Thorne:2017jft,Bartolo:2017szm,Bartolo:2018elp} (for example breaking the tensor consistency relation \citep{Duivenvoorden:2019ses}). Diagnosing the statistical properties of the tensor sector is thus a powerful way to probe the GW origins and shed light on inflationary physics; furthermore, we note that large GW bispectra are not \textit{a priori} ruled out by the current bounds on tensor power spectra \citep[e.g.,][]{Namba:2015gja,Dimastrogiovanni:2016fuu,Gong:2023kpe,Ozsoy:2019slf,Mylova:2019jrj,Cabass:2021fnw,Cabass:2022jda,Bordin:2020eui,Cabass:2021iii,Shiraishi:2011st,Bartolo:2018elp}. Motivated by this, we here perform an analysis to jointly test the parity and Gaussianity of primordial GWs (or gravitons) employing three-point correlators (equivalently bispectra) of the CMB temperature and polarization fields. This builds upon previous work including both a range of theoretical GW bispectrum predictions \citep{Shiraishi:2011st,Shiraishi:2012sn,Shiraishi:2013kxa,Namba:2015gja,Shiraishi:2016yun,Bartolo:2018elp}, and data analyses involving WMAP temperature ($T$) data \citep{Shiraishi:2014ila} and {\it Planck} temperature and $E$-mode polarization ($T$ and $E$) \citep{Planck:2015zfm,Planck:2019kim}, none of which have yet found any significant deviation from parity invariance and Gaussianity (see \citep{Shiraishi:2019yux} for review).%
\footnote{
For the scalar sector, parity-violation appears first as an imaginary component of the 4-point functions (in the limit of negligible isotropy violation \citep{Shiraishi:2016mok}).
%the imaginary signal of 4-point correlator becomes the lowest-order indicator of parity violation because observed isotropy violation is so small \citep{Shiraishi:2016mok}}. 
See \citep{Philcox:2021hbm,Cahn:2021ltp,Hou:2022wfj,Philcox:2022hkh,Cabass:2022oap,CyrilCS,PhilcoxCMB,Philcox:2023ypl} for searches for this signal.}

In this paper, we, for the first time,\footnote{\citep{Coulton:2019bnz} also analyzed the $B$-mode polarization data, but only in the context of galactic foregrounds. Furthermore, \citep{Shiraishi:2014roa,Shiraishi:2019exr,Duivenvoorden:2019ses} present a detailed discussion of $B$-mode bispectra in the context of template analyses, but do not apply their methodology to data.} include the \textit{Planck} $B$-mode polarization ($B$) field in the data analysis (which dramatically enhances constraints on the parity-odd sector) and consider both general tests for non-Gaussianity, and constraints on a specific inflationary template. This is facilitated both by new data (\textit{Planck} PR4) with better control of large-scale polarization systematics, and a suite of new CMB polyspectral estimators \citep{Philcox:2023uwe,Philcox:2023psd,PolyBin}, which allow for quasi-optimal estimation of binned parity-even and parity-odd $T$, $E$, and $B$-mode bispectra, whilst simultaneously accounting for observational effects such as mask-induced leakage between bins and polarizations. These have been previously used to assess parity-violation in the scalar CMB sector for both temperature and polarization \citep{PhilcoxCMB,Philcox:2023ypl}.

Parity-even and parity-odd information on graviton non-Gaussianity can be easily distinguished by working in harmonic space, where, for the CMB bispectrum, there is a bifurcation into even and odd $\ell_1 + \ell_2 + \ell_3$ \citep{Kamionkowski:2010rb,Shiraishi:2011st,Shiraishi:2014roa}. Using the aforementioned recently developed general bispectrum estimator \citep{Philcox:2023uwe,Philcox:2023psd,PolyBin} (building on previous binned estimators \citep{Bucher:2015ura,Bucher:2009nm} and numerical tricks \citep{2011MNRAS.417....2S,2015arXiv150200635S}), we separately extract the parity-even and parity-odd information from CMB bispectrum datasets and perform model-independent tests of parity violation using the GW three-point function, essentially asking whether we detect any signal above the noise. Following this, we place constraints on the amplitude of a well-examined GW bispectrum template \citep{Shiraishi:2013kxa,Planck:2015zfm,Shiraishi:2019yux}, parametrized by the non-Gaussianity parameter $f_{\rm NL}^{ttt}$ (see \eqref{eq: hhh_fNLttt} for a concrete form). This template can be defined for both the parity-even and parity-odd sector and has a dominant signal for equilateral momentum configurations $k_1 \sim k_2 \sim k_3$. As such, it can well describe non-Gaussian chiral GWs sourced by gauge fields coupled to axions in various ways \citep{Cook:2013xea,Namba:2015gja,Agrawal:2017awz}. We place constraints on $f_{\rm NL}^{ttt}$ from both parity-even and -odd information alone as well as the combined (maximally chiral) limit; our results are consistent with zero apart from $\simeq 2.5\sigma$ parity-even signal extracted from $E$-modes or their combination with $B$-modes. The most stringent value is obtained from the full $T$, $E$ and $B$ dataset, with $f_{\rm NL}^{ttt} = 900 \pm 700$, which represents an improvement over previous constraints ($f_{\rm NL}^{ttt} = 1300 \pm 1200$ \citep{Planck:2019kim}) by a factor greater than two, mostly owing to the inclusion of $B$-modes.\footnote{
This $f_{\rm NL}^{ttt}$ is equivalent to $f_{\rm NL}^P$ in ~\citep{Shiraishi:2014ila}, $f_{\rm NL}^{\rm tens}$ in \citep{Planck:2015zfm,Planck:2019kim} and $f_{\rm NL}^{ttt, \rm eq}$ in \citep{Shiraishi:2019yux}.
}

The remainder of this paper is structured as follows. In \S\ref{sec: theory-model} we present theoretical setup, as well as our motivations for testing parity and Gaussianity of the graviton sector. \S\ref{sec: data} discusses the \textit{Planck} data as well as various aspects of its analysis, including bispectrum estimation and simulations. In \S\ref{sec: results}, we present our main result: constraints on the tensor amplitude parameter $f_{\rm NL}^{ttt}$, before concluding in \S\ref{sec: summary}. Throughout this work, we assume a \textit{Planck} 2018 fiducial cosmology with parameter set $\{\omega_b = 0.022383, \omega_{cdm} = 0.12011, h = 0.6732, \tau_{\rm reio} = 0.0543, \log 10^{10}A_s = 3.0448, n_s = 0.96605\}$ for a single massive neutrino with $m_\nu = 0.06\,\mathrm{eV}$ \citep{2020A&A...641A...6P}. To avoid confirmation bias, the bispectrum estimation and analysis pipeline was finalized before the \textit{Planck} bispectrum was computed.

\section{Theory}\label{sec: theory-model}

\noindent
We begin by outlining the theoretical fundamentals of graviton non-Gaussianity, connecting them to CMB statistics, and discussing the specific CMB template whose amplitude will be constrained in \S\ref{subsec: results-model}. \resub{Here and henceforth, we will ignore correlations between the scalar and tensor sectors; the corresponding mixed non-Gaussianity represents an interesting extension of the ideas considered in this work.}

\subsection{Primordial Tensors}
\noindent In the synchronous gauge, we may write the metric perturbation induced by a graviton as $h_{ij} \equiv \delta g_{ij}^{\rm T} / a^2$, where $a$ is the scale factor and $\delta g_{\mu\nu}^{\rm T}$ is the tensor part of the metric perturbation (at linear order). This can be decomposed into helicity-$\pm2$ states (equivalent to circular $R/L$ polarizations) via
\beq
h_{ij}(\vx) =
\int_\vk
\sum_{\lambda = \pm 2} 
h_\vk^{(\lambda)}
e_{ij}^{(\lambda)}(\hk)
 e^{i \vk \cdot \vx} ,
 \eeq
for transverse-traceless helicity state $e_{ij}^{(\lambda)}(\hk)$, obeying 
 \beq
 	e_{ij}^{(\lambda)}(-\hk) = e_{ij}^{(-\lambda)}(\hk) = e_{ij}^{(\lambda) *}(\hk), \qquad e_{ij}^{(\lambda)}(\hk) e^{(\lambda'),ij}(\hk) = 2 \delta_{\rm K}^{\lambda, -\lambda'}
\eeq
\citep{Shiraishi:2010kd}, with $\int_{\vk}\equiv (2\pi)^{-3}\int d^3 \vk$ throughout. The mode functions $h_{\vk}^{(\lambda)}$ set the statistical properties of the graviton, with the three-point function given by
\beq\label{eq: tensor-bispectrum}
	\delD{\vk_1+\vk_2+\vk_3}B_{h}^{(\lambda_1\lambda_2\lambda_3)}(\vk_1,\vk_2,\vk_3)\equiv \left\langle \prod_{n=1}^3 h_{\vk_n}^{(\lambda_n)}\right\rangle
\eeq
Under a conjugation, $h_{\vk}^{(\lambda)}\to h_\vk^{(\lambda)*}\equiv h_{-\vk}^{(\lambda)}$, thus 
\beq
	B_{h}^{(\lambda_1\lambda_2\lambda_3)*}(\vk_1,\vk_2,\vk_3) = B_{h}^{(\lambda_1\lambda_2\lambda_3)}(-\vk_1,-\vk_2,-\vk_3).
\eeq
A parity transform sends $\vx \to \mathbb{P}[\vx] \equiv - \vx$; in this case, the helicity-state transforms as $\mathbb{P}\left[h_\vk^{(\lambda)}\right] \to h_{- \vk}^{(-\lambda)}$, and 
\beq\label{eq: parity-sym}
	\mathbb{P}\left[B_{h}^{(\lambda_1\lambda_2\lambda_3)}(\vk_1,\vk_2,\vk_3)\right]= B_{h}^{(-\lambda_1-\lambda_2-\lambda_3)}(-\vk_1,-\vk_2,-\vk_3)\equiv \left[B_{h}^{(-\lambda_1-\lambda_2-\lambda_3)}(\vk_1,\vk_2,\vk_3)\right]^*.
\eeq
We can further decompose the bispectrum into parity-even/odd components as $B_h^{(\lambda_1\lambda_2\lambda_3)}(\vk_1,\vk_2,\vk_3) = B_h^{(\lambda_1\lambda_2\lambda_3),+}(\vk_1,\vk_2,\vk_3) + B_h^{(\lambda_1\lambda_2\lambda_3),-}(\vk_1,\vk_2,\vk_3)$, which satisfy
\beq\label{eq: bispec-parity}
	\mathbb{P}\left[B_{h}^{(\lambda_1\lambda_2\lambda_3),\pm}(\vk_1,\vk_2,\vk_3)\right] = \pm\,B_{h}^{(\lambda_1\lambda_2\lambda_3),\pm}(\vk_1,\vk_2,\vk_3);
\eeq
from \eqref{eq: parity-sym}, these have the symmetry $B_{h}^{(\lambda_1\lambda_2\lambda_3),\pm}(\vk_1,\vk_2,\vk_3) = \pm\,B_{h}^{(-\lambda_1-\lambda_2-\lambda_3),\pm}(-\vk_1,-\vk_2,-\vk_3)$.%from the above properties, $B_h^{(\lambda_1\lambda_2\lambda_3),\pm}+B_h^{(-\lambda_1-\lambda_2-\lambda_3),\pm}$ is purely real / imaginary.

\subsection{CMB Fluctuations}

%Denoting the parity-even/odd part of the GW bispectrum by $\left\langle \prod_{n=1}^3 h_{\vk_n}^{(\lambda_n)}\right\rangle^{+/-}$, obeys the relation
 %\beq \label{eq: hhh_parity}
 %\left\langle \prod_{n=1}^3 h_{\vk_n}^{(\lambda_n)}  \right\rangle^\pm  = \pm  \left\langle \prod_{n=1}^3 h_{- \vk_n}^{(- \lambda_n)}  \right\rangle^\pm .
 %\eeq

\noindent Large-scale tensor-modes source both CMB temperature and polarization fluctuations, which are here denoted by the harmonic coefficients $a_{\ell m}^X$ with $X\in\{T,E,B\}$. At linear order,  
\beq\label{eq: tensor-transfer-alm}
a_{\ell m}^{X} &=& 
4\pi i^{\ell} \int_{\vk}
{\cal T}_{\ell}^{X}(k)
\left[ h_\vk^{(+2)} {}_{-2} Y_{\ell m}^*(\hk)
  + (-1)^x h_\vk^{(-2)} {}_{+2} Y_{\ell m}^*(\hk) \right] \nonumber \\
  &\equiv& 4\pi i^{\ell} \int_\vk
{\cal T}_{\ell}^{X}(k)
\left[ h_\vk^{(+2)} + (-1)^{x + \ell} h_{-\vk}^{(-2)} \right] {}_{-2} Y_{\ell m}^*(\hk)
\eeq
\citep{Shiraishi:2010sm,Shiraishi:2010kd}, where ${\cal T}_{\ell}^{X}$ is the tensor-mode linear transfer function, ${}_{s}Y_{\ell m}$ is a spin-weighted spherical harmonic $x$ encodes the parity of the field; $0$ for $T,E$ and $1$ for $B$. 

The angular bispectrum follows directly from \eqref{eq: tensor-transfer-alm} and \eqref{eq: tensor-bispectrum}
\beq
	\av{\prod_{n=1}^3a_{\ell_nm_n}^{X_n}} &=& \prod_{n=1}^3\left[4\pi i^{\ell_n} \int_{\vk_n}{\cal T}_{\ell_n}^{X_n}(k_n){}_{-2}Y_{\ell_nm_n}^*(\hk_n)\right]\delD{\vk_1+\vk_2+\vk_3}\\\nonumber
	&&\times \bigg\{B^{(222)}_h(\vk_1,\vk_2,\vk_3)+(-1)^{\ell_{123}+x_{123}}B_h^{(-2-2-2)}(-\vk_1,-\vk_2,-\vk_3)\\\nonumber
	&&\qquad\,+\,(-1)^{x_1+\ell_1}B^{(-222)}_h(-\vk_1,\vk_2,\vk_3)+(-1)^{x_{23}+\ell_{23}}B^{(2-2-2)}_h(\vk_1,-\vk_2,-\vk_3)\\\nonumber
	&&\qquad\,+\,(-1)^{x_2+\ell_2}B^{(2-22)}_h(\vk_1,-\vk_2,\vk_3)+(-1)^{x_{13}+\ell_{13}}B^{(-22-2)}_h(-\vk_1,\vk_2,-\vk_3)\\\nonumber
	&&\qquad\,+\,(-1)^{x_3+\ell_3}B^{(22-2)}_h(\vk_1,\vk_2,-\vk_3)+(-1)^{x_{12}+\ell_{12}}B^{(-2-22)}_h(-\vk_1,-\vk_2,\vk_3)\bigg\}
\eeq
for $\ell_{ab}\equiv \ell_a+\ell_b$ \textit{et cetera}. Splitting into parity-even and odd graviton bispectra via \eqref{eq: bispec-parity}, this simplifies considerably
\beq\label{eq: cmb-bispectra}
	\av{\prod_{n=1}^3a_{\ell_nm_n}^{X_n}} &=& \sum_\pm\left[1\pm (-1)^{\ell_{123}+x_{123}}\right]\prod_{n=1}^3\left[4\pi i^{\ell_n} \int_{\vk_n}{\cal T}_{\ell_n}^{X_n}(k_n){}_{-2}Y_{\ell_nm_n}^*(\hk_n)\right]\delD{\vk_1+\vk_2+\vk_3}\\\nonumber
	&&\times \bigg\{B^{(222),\pm}_h(\vk_1,\vk_2,\vk_3)+(-1)^{x_1+\ell_1}B^{(-222),\pm}_h(-\vk_1,\vk_2,\vk_3)\\\nonumber
	&&\qquad\,+\,(-1)^{x_2+\ell_2}B^{(2-22),\pm}_h(\vk_1,-\vk_2,\vk_3)+(-1)^{x_3+\ell_3}B^{(22-2),\pm}_h(\vk_1,\vk_2,-\vk_3)\bigg\}
\eeq

%The angular bispectrum  is hence written down as
%\beq
%\left\langle \prod_{n=1}^3 a_{\ell_n m_n}^{X_n} \right\rangle
%&=& \left[ \prod_{n=1}^3 4\pi i^{\ell_n} \int \frac{d^3 k_n}{(2\pi)^{3}}
%  {\cal T}_{\ell_n}^{X_n}(k_n) {}_{-2} Y_{\ell_n m_n}^*(\hk_n) \right] \nonumber \\
%&& \times \left[ \left\langle h_{\vk_1}^{(+2)} h_{\vk_2}^{(+2)} h_{\vk_3}^{(+2)} \right\rangle + (-1)^{\sum_{n=1}^3 (x_n  + \ell_n)} \left\langle h_{-\vk_1}^{(-2)} h_{-\vk_2}^{(-2)} h_{-\vk_3}^{(-2)} \right\rangle \right. \nonumber \\
%  %@@@
%  && \left.\quad + (-1)^{x_3 + \ell_3} \left( \left\langle h_{\vk_1}^{(+2)} h_{\vk_2}^{(+2)} h_{- \vk_3}^{(-2)} \right\rangle + (-1)^{\sum_{n=1}^3 (x_n  + \ell_n)} \left\langle h_{-\vk_1}^{(-2)} h_{-\vk_2}^{(-2)} h_{\vk_3}^{(+2)} \right\rangle \right) \right. \nonumber \\
%  %@@@
%&& \left.\quad + (-1)^{x_2 + \ell_2} \left( \left\langle h_{\vk_1}^{(+2)} h_{-\vk_2}^{(-2)} h_{\vk_3}^{(+2)} \right\rangle + (-1)^{\sum_{n=1}^3 (x_n  + \ell_n)} \left\langle h_{-\vk_1}^{(-2)} h_{\vk_2}^{(+2)} h_{-\vk_3}^{(-2)} \right\rangle \right) \right. \nonumber \\
%  %@@@
%&& \left.\quad + (-1)^{x_1 + \ell_1} \left( \left\langle h_{-\vk_1}^{(-2)} h_{\vk_2}^{(+2)} h_{\vk_3}^{(+2)} \right\rangle + (-1)^{\sum_{n=1}^3 (x_n  + \ell_n)} \left\langle h_{\vk_1}^{(+2)} h_{-\vk_2}^{(-2)} h_{-\vk_3}^{(-2)} \right\rangle \right)
%  \right] .
%\eeq
From the prefactor it is evident that parity-even (odd) GW bispectra $B_h^{\pm}$ source non-vanishing CMB bispectrum signal only with $\ell_1 + \ell_2 + \ell_3 + x_1 + x_2 + x_3 = \rm even\,(odd)$ \citep{Shiraishi:2011st,Shiraishi:2016ads}. In other words, 
\begin{itemize}
\item Parity-even GW bispectra $B_h^+$ source non-zero CMB bispectra with even $\ell_1+\ell_2+\ell_3$ if there are zero or two $B$-modes, and odd $\ell_1+\ell_2+\ell_3$ else.
\item Parity-odd GW bispectra $B_h^-$ source non-zero CMB bispectra with odd $\ell_1+\ell_2+\ell_3$ if there are zero or two $B$-modes, and even $\ell_1+\ell_2+\ell_3$ else.
\end{itemize}
This delineation is slightly more complex than in previous analyses \citep{Shiraishi:2014ila,Planck:2015zfm,Planck:2019kim}, due to the presence of parity-odd $B$-modes. In \S\ref{sec: data}, we will will measure the parity-even and parity-odd signals separately on the basis of this rule, and test the parity symmetry of the graviton bispectrum. A non-zero signal in a parity-odd correlator thus would give evidence of parity-violating non-Gaussianity in the graviton sector.

\subsection{A Well-Motivated Template}\label{subsec: theory-template}
\noindent To perform model-specific CMB GW analyses, one need to specify the form of the graviton three-point function $B_h^{(\lambda_1\lambda_2\lambda_3)}$. In this work, we consider a commonly used phenomenological template specified by
\beq \label{eq: hhh_fNLttt}
B_h^{(\lambda_1\lambda_2\lambda_3)}(\vk_1,\vk_2,\vk_3) = \frac{16\sqrt{2}}{27} f_{\rm NL}^{ttt}\,\times\,B^{\rm eq}_\zeta(k_1,k_2, k_3)
  e_{ij}^{(-\lambda_1)}(\hk_1) e_{jk}^{(-\lambda_2)}(\hk_2) e_{ki}^{(-\lambda_3)}(\hk_3)  
 \delta_{\lambda_1, +2}^{\rm K} \delta_{\lambda_2, +2}^{\rm K} \delta_{\lambda_3, +2}^{\rm K} , 
  \eeq
\citep{Shiraishi:2013kxa,Shiraishi:2014ila,Planck:2015zfm,Planck:2019kim,Shiraishi:2019yux}, where $B^{\rm eq}_\zeta$ is the standard equilateral-type scalar bispectrum, defined by
\beq
  B^{\rm eq}_\zeta(k_1,k_2, k_3) \equiv \frac{18}{5} (2\pi^2 A_s)^2
  \left[
    - \left( \frac{1}{k_1^3 k_2^3} + 2 \, {\rm perms.}  \right)
    - \frac{2}{k_1^2 k_2^2 k_3^2} 
    + \left( \frac{1}{k_1 k_2^2 k_3^3} + 5 \, {\rm perms.}  \right)
    \right] .
  \eeq
assuming scale-invariance. The bispectrum template of \eqref{eq: hhh_fNLttt} contains only positive helicity modes; as such, it is maximally chiral, containing both parity-even and parity-odd components with the same primordial amplitude, such that $B_h^{(\lambda_1\lambda_2\lambda_3),+}(\vk_1,\vk_2,\vk_3) = B_h^{(\lambda_1\lambda_2\lambda_3),-}(\vk_1,\vk_2,\vk_3) = 
(1/2) B_h^{(\lambda_1\lambda_2\lambda_3)}(\vk_1,\vk_2,\vk_3)$.\footnote{This can be split into distinct parity-even and parity-odd components through \eqref{eq: bispec-parity}, which can be analyzed separately.} A dominant signal arises in the equilateral configurations $k_1 \simeq k_2 \simeq k_3$. \resub{Notably, this is just one choice of template; one could also other forms such as local-type tensor non-Gaussianity (sourced, for example, by magnetic fields) or that induced by Weyl gravity \citep{Shiraishi:2011st,Shiraishi:2012sn,Shiraishi:2019yux}.} 

The GW bispectra encoded by the above template can be produced in inflationary models that involve couplings between axions ($\chi$) and gauge fields ($A$) through the Lagrangian
\beq
    {\cal L}(A_\mu,\chi) \supset - \frac{1}{4} F_{\mu\nu}F^{\mu\nu} - \frac{\chi}{4f} F_{\mu\nu}\tilde{F}^{\mu\nu}
\eeq
\citep{Sorbo:2011rz,Barnaby:2011vw} (displayed for $U(1)$ gauge fields, though higher-order symmetries are also possible). Here, $F_{\mu\nu}\equiv 2\,\partial_{[\mu}A_{\nu]}$ is the gauge field strength, $\tilde{F}_{\mu\nu}$ is its dual, and $f$ is the Chern-Simons coupling parameter \citep{Sorbo:2011rz,Barnaby:2011vw,Barnaby:2012xt,Cook:2013xea,Namba:2015gja,Dimastrogiovanni:2016fuu,Agrawal:2017awz,Watanabe:2020ctz,Mirzagholi:2020irt}. In such models, one of the two gauge field helicity states experiences a tachionic instability and is explosively produced for Fourier modes with physical wavelengths comparable to the horizon. Through the gravitational interaction, the gauge field quadratically sources metric perturbations (with $h\sim A^2$); thus, supposing Gaussianity of $A_\mu$, the generated GW becomes chi-squared distributed and thus highly non-Gaussian field. In this process, the chirality of the gauge field is transmitted to the GW. Without loss of generality, we identify the growing mode with $h^{(+2)}$, allowing us drop any correlators involving $h^{(-2)}$.

Due to the gauge field production for limited Fourier modes, the chiral GW bispectrum has a dominant signal when the three bispectrum wavenumbers are approximately equal, \textit{i.e}\ it peaks in equilateral configurations with $k_1 \simeq k_2 \simeq k_3$.\footnote{If one instead considers a Lagrangian of the form ${\cal L} \supset f(\chi) \left( - \frac{1}{4} F^2 + \frac{\gamma}{4} F \tilde{F} \right)$ and appropriately fine-tunes the time evolution of the coupling function $f(\chi)$, one can produce a gauge field that does not decay on superhorizon scales. In this case, a sizable bispectrum signal can then arise in the squeezed limit: $k_1 \sim k_2 \gg k_3$, $k_2 \sim k_3 \gg k_1$ and $k_3 \sim k_1 \gg k_2$ \citep{Bartolo:2015dga}.} Moreover, if the axion rolls slowly down a not-so-steep potential for a not-so-small period, the GW bispectrum can be shown to take the nearly scale-invariant shape of \eqref{eq: hhh_fNLttt}.\footnote{See \citep{Namba:2015gja,Dimastrogiovanni:2016fuu} for cases inducing signals with strong scale-dependence.} In cases where the gauge field has $U(1)$ symmetry, the tensor non-linearity parameter is evaluated as \beq
    \left.f_{\rm NL}^{ttt}\right|_{U(1)} \simeq 6.4 \times 10^{11} A_s^3 \epsilon^3 \frac{e^{6 \pi \xi}}{\xi^9},
\eeq
where $\epsilon$ is the inflaton slow-roll parameter and $\xi \equiv \dot{\chi}/(2fH)$, with $\dot{\chi}$ being the time derivative of the axion field and $H$ the Hubble parameter \citep{Cook:2013xea,Planck:2015zfm}. In models involving an $SU(2)$ gauge field, the size of non-Gaussianity is determined by the energy density fraction of the gauge field $\Omega_A$ and the tensor-to-scalar ratio $r$ as \citep{Agrawal:2017awz}
\beq
    \left.f_{\rm NL}^{ttt}\right|_{SU(2)} \simeq 2.5 \frac{r^2}{\Omega_A}.
\eeq 

In \S\ref{subsec: results-model}, we estimate $f_{\rm NL}^{ttt}$ from \textit{Planck} $T, E$ and $B$-modes by inserting the above template in the bispectrum relation of \eqref{eq: cmb-bispectra}. We will give results both for the parity-even/odd signal (denoting the respective amplitudes as $f_{\rm NL}^{ttt, +/-}$) and the combined one (with amplitude $f_{\rm NL}^{ttt} \equiv f_{\rm NL}^{ttt,+}+f_{\rm NL}^{ttt,-}$). For this purpose, we may restrict the range of CMB multipoles to $\ell \in [2, 500)$ as we expect no gain in the signal-to-noise ratio when including $\ell \gtrsim 500$ due to the rapid damping of CMB tensor modes \citep{Shiraishi:2013kxa,Shiraishi:2019yux}.

\section{Dataset \& Analysis Pipeline}\label{sec: data}
\subsection{Observational Data}
\noindent In this work, we obtain constraints on gravitational wave bispectra using the latest temperature and polarization maps provided by \textit{Planck}. Our dataset is analogous to that used in \citep{Philcox:2023ypl}; we use the PR4 dataset, consisting of component-separated temperature and $Q/U$-mode polarization maps processed using the \textsc{npipe} pipeline in combination with the \textsc{sevem} component separation algorithm \citep{Planck:2020olo,Rosenberg:2022sdy,Tristram:2020wbi}.\footnote{Publicly available at \href{https://portal.nersc.gov/project/cmb/planck2020/}{portal.nersc.gov/project/cmb/planck2020}.} These are an updated version of the \textit{Planck} 2018 dataset (PR3), and have a number of improvements, in particular with regards to the treatment of large-scale polarization data and reduction of systematics. We additionally utilize 600 FFP10 simulations (all processed with \textsc{npipe}); 100 will be used to construct the quasi-optimal bispectrum estimators and the remainder will be used for validation and to estimate the variance of the observed statistics \citep{Planck:2020olo}. 

Following \citep{Planck:2019kim}, high-emission regions of the galactic plane are removed via the \textit{Planck} common component-separation mask (smoothed at ten arcminute scales), which preserves $\simeq 80\%$ of the sky \citep{Planck:2018yye}. In the analyses below, we additionally apply the \textit{Planck} beam, which includes the PR4 polarization transfer function on large scales ($\ell<40$), extracted from the FFP10 simulations \citep{Planck:2020olo}. Since the gravitational wave signatures we consider in this study are predominantly confined to large angular scales, we restrict our attention to $\ell_{\rm max}=500$, choosing a conservative \textsc{healpix} pixellation strategy with $N_{\rm side}=512$ \citep{Gorski:2004by} for all maps. 

Before computing bispectra (\S\ref{subsec: bispectra}), all maps are filtered by a linear operator, $\Si$ \citep[cf.][]{2011MNRAS.417....2S,2015arXiv150200635S}. As discussed in \citep{Philcox:2023uwe,Philcox:2023psd}, the choice of $\Si$ sets the precision of the output CMB correlators, with optimal variances achieved if $\Si$ is equal to the true inverse pixel covariance of the dataset. In practice, such a choice is difficult to implement, thus we here adopt a simpler scheme following \citep{Philcox:2023ypl}. Introducing a mask $W(\hn)$ and a linear inpainting operator $\Pi$, the action of $\Si$ on some map $a^X(\hn)$ (for $X\in \{T,E,B\}$) can be written
\beq
    [\Si a]^{X}_{\ell m} = \sum_Y S^{-1,XY}_{\ell}\int d\hn\,Y_{\ell m}^*(\hn)W(\hn)\Pi[a^Y](\hn);
\eeq
\textit{i.e.}\ we inpaint small holes in the map \citep[cf.][]{Gruetjen:2015sta}, mask the galactic plane, transform to harmonic space, then normalize by a rotationally-invariant filter. Here, we fix $S_\ell^{XY} = C_\ell^{XY}+\delta_{\rm K}^{XY}N_\ell^{X}$ with fiducial CMB power spectrum $C_\ell^{XY}$ and diagonal noise $N_\ell^X$, as measured from the high-resolution \textit{Planck} half-mission maps. We stress that our choice of $\Si$ filter cannot induce bias; it is fully accounted for in the bispectrum estimator discussed below, and any simplification just leads to a reduction in sensitivity.

\subsection{Bispectrum Estimation}\label{subsec: bispectra}
\noindent In the remainder of this work, we will consider only \textit{reduced bispectra} $b_{\ell_1\ell_2\ell_3}^{XYZ}$, which are defined from CMB maps $a_{\ell m}^X$ via
\beq\label{eq: reduced-bis-def}
    \av{a_{\ell_1m_1}^Xa_{\ell_2m_2}^Ya_{\ell_3m_3}^Z}_c &=&\frac{1}{3}\sqrt{\frac{(2\ell_1+1)(2\ell_2+1)(2\ell_3+1)}{4\pi}}\tj{\ell_1}{\ell_2}{\ell_3}{m_1}{m_2}{m_3}\left[\tj{\ell_1}{\ell_2}{\ell_3}{-1}{-1}{2}+\text{2 perms.}\right]b^{XYZ}_{\ell_1\ell_2\ell_3}\\\nonumber
    &\equiv& w^{\ell_1\ell_2\ell_3}_{m_1m_2m_3}b^{XYZ}_{\ell_1\ell_2\ell_3},
\eeq
where the weights $w^{\ell_1\ell_2\ell_3}_{m_1m_2m_3}$ are a generalization of the Gaunt integral, allowing for odd $\ell_1+\ell_2+\ell_3$ \citep{Shiraishi:2014roa,Philcox:2023psd}. This restricts to rotationally invariant bispectra (due to the first Wigner $3j$ symbol), reducing the dimensionality from $6$ to $3$ angular variables. Conjugation symmetries of $a_{\ell m}^X$ imply that bispectra with even (odd) $\ell_1+\ell_2+\ell_3$ are real (imaginary).

To estimate bispectra, we use the \href{https://github.com/oliverphilcox/PolyBin}{\textsc{PolyBin}} code\footnote{Available at \href{https://github.com/oliverphilcox/PolyBin}{GitHub.com/OliverPhilcox/PolyBin} \citep{PolyBin}.} described in \citep{Philcox:2023uwe,Philcox:2023psd} (which builds upon template-based techniques \citep[e.g.,][]{2011MNRAS.417....2S,2015arXiv150200635S}, the binned estimators of \citep{Bucher:2009nm,Bucher:2015ura}, and the three-dimensional estimators of \citep{2021PhRvD.103j3504P,Philcox:2021ukg}). Rather than computing the full bispectrum (which has dimensionality $\mathcal{O}(\ell_{\rm max}^3)$), this directly estimates the bispectrum in some set of angular bins, $\vec b\equiv \{b_1,b_2,b_3\}$, hereafter denoted $b^{XYZ}_\chi(\vec b)$, where $\chi=\pm 1$ indicates the parity of the correlator. In more detail:
\begin{itemize}
    \item \resub{\textbf{Parity-even , $\chi=+1$}: this corresponds to \textit{even} $\ell_1+\ell_2+\ell_3$ and \textit{real} bispectra for correlators with zero or two $B$-modes, and \textit{odd} $\ell_1+\ell_2+\ell_3$ and \textit{imaginary} bispectra else}.
    \item \resub{\textbf{Parity-odd , $\chi=-1$}: this corresponds to \textit{odd} $\ell_1+\ell_2+\ell_3$ and \textit{imaginary} bispectra for correlators with zero or two $B$-modes, and \textit{even} $\ell_1+\ell_2+\ell_3$ and \textit{real} bispectra else}.
\end{itemize}
\textit{Viz} the discussion in \S\ref{subsec: theory-template}, parity-even (parity-odd) inflationary templates such as $f_{\rm NL}^{ttt,+}$ ($f_{\rm NL}^{ttt,-}$) will correspond to $\chi=+1$ ($\chi=-1$).

Schematically, \textsc{PolyBin} estimates bispectra using two components: (1) a numerator, $[\mathcal{F}\widehat{b}]$, involving three copies of the data; (2) a data-independent normalization matrix, $\mathcal{F}$ (often called the Fisher matrix).\footnote{When forming the estimator one has freedom in where to include the mask, $W(\hn)$. Following \citep{PhilcoxCMB,Philcox:2023ypl}, we here include it in the $\Si$ weights rather than directly in the estimator (as in \citep{Philcox:2023uwe,Philcox:2023psd}) since this (empirically) reduces bin-to-bin leakage without inducing bias.} As discussed in \citep{Philcox:2023uwe,Philcox:2023psd}, this is an unbiased estimator for any choice of weighting scheme, implying that the output spectra can be compared to theory without the need to account for window function convolution or leakage between different bins, parities, and fields. Furthermore, in the ideal limit, where $\Si$ is equal to the inverse pixel covariance and $a_{\ell m}^X$ is Gaussian, the estimator is optimal and has covariance $\mathcal{F}^{-1}$ \citep[e.g.,][]{Hamilton:1999uw}. 

Given beam $B_\ell^X$ and triplet of fields $u\equiv \{XYZ\}$ containing $n_B$ $B$-modes, the estimator numerator takes the schematic form
\beq\label{eq: rough-bispectrum-estimator}
    \left[\mathcal{F}\widehat{b}\right]^{u}_{\chi}(\vec b) &\sim& \sum_{\ell_i\in b_i}\sum_{m_i}\left[1+\chi(-1)^{\ell_1+\ell_2+\ell_3+n_B}\right]B_{\ell_1}^{u_1}B_{\ell_2}^{u_2}B_{\ell_3}^{u_3}w^{\ell_1\ell_2\ell_3}_{m_1m_2m_3}\left[a_{\ell_1m_1}^{u_1}a_{\ell_1m_1}^{u_1}a_{\ell_1m_1}^{u_1}-3\av{a_{\ell_1m_1}^{u_1}a_{\ell_2m_2}^{u_2}}a_{\ell_3m_3}^{u_3}\right],
\eeq
summing over all $\ell$ modes in the bin of interest. This involves both a cubic and a linear term; the latter does not change the mean of the estimator, but can considerably reduce the variance on large-scales. The normalization matrix has a more complex form, which is set by demanding that the estimator be unbiased (such that $\mathbb{E}[\widehat{b}_\chi^u(\vec b)] = b_\chi^u(\vec b)$ for expectation operator $\mathbb{E}$). In practice, \textsc{PolyBin} does not compute the $\mathcal{O}(\ell_{\rm max}^6)$ summations in the above expressions directly: efficient use of spherical harmonic transforms and Monte Carlo summation allows both the numerator and denominate to be efficiently computed with $\mathcal{O}(\ell_{\rm max}^3)$ complexity. All computations are done in \textsc{Python}, with spherical harmonic transforms performed using \textsc{libsharp} \citep{2013A&A...554A.112R}.

In this work, we compute bispectra for each of the ten non-trivial combinations of the $T, E, B$ fields: 
\beq\label{eq: field-triplets}
\{TTT, TTE, TTB, TEE, TEB, TBB, EEE, EEB, EBB, BBB\}
\eeq
with $\chi\in\{\pm 1\}$. Based on the discussion in \S\ref{subsec: theory-template}, we restrict to $\ell_{\rm max}=500$; noting that the computation time of the estimator denominator is cubic in the number of $\ell$-bins, $N_\ell$ (but linear in the total number of bins, $N_{\rm bins}$), we use $N_\ell = 13$ $\ell$-bins roughly equally spaced in $\ell^{2/3}$ (ensuring roughly equal signal-to-noise per bin), with bin-edges $\{2,   3,   4,  22,  49,  82, 121, 164, 211, 262, 317, 375, 436, 500\}$. This gives a total of $N_{\rm bins} = 5972$ bispectrum bins (dropping any empty bins and removing degeneracies by restricting to $b_1\leq b_2\leq b_3$ for $TTT$ \textit{et cetera}).\footnote{In the analysis of \S\ref{sec: results}, we drop $5$ of these bins, since their variance is both extremely large and poorly measured.} To form the linear term of the bispectrum estimator \eqref{eq: rough-bispectrum-estimator}, we require the expectation $\av{a_{\ell_1m_1}^{u_1}a_{\ell_2m_2}^{u_2}}$; rather than computing this analytically (which is expensive, since the mask breaks rotational invariance), we estimate the entire linear term as a Monte Carlo average of the data and $N_{\rm sim}=100$ FFP10 simulations (requiring $\mathcal{O}(N_{\rm sim}N_{\ell})$ computational costs, if harmonic transforms are rate-limiting), as discussed in \citep{Philcox:2023psd}. Furthermore, the normalization matrix is also computed as a Monte Carlo average; due to the efficient numerical schemes introduced in \citep{2011MNRAS.417....2S}, we are able to estimate the matrix to percent-level precision using only $N_{\rm mc}=10$ realizations (with $\mathcal{O}(N_{\rm mc}N_{\rm bin})$ computational cost).

Using the above hyperparameters, computing the bispectrum numerator of a single dataset required $30$ CPU-hours and around $8000$ harmonic transforms (with $25\,\mathrm{GB}$ memory). This is dominated by the linear term; if this is removed (which we later show to yield only a minor inflation in our constraints), the computation time reduces to only $5$ CPU-minutes with $80$ transforms. The normalization matrix (which is independent of the data) required $30$ CPU-hours and $60\,000$ harmonic transforms per realization on a (partially utilized) high-performance node with $3\,\mathrm{TB}$ memory. In total, full computation of the bispectra of 500 FFP10 simulations and the \textit{Planck} data required around $20\,000$ CPU-hours. 

\subsection{Binning the Theoretical Templates}
\noindent To compare the data with bispectrum templates (such as \eqref{eq: hhh_fNLttt}), one must recast the theory into the same form as the data. As discussed in \citep{Philcox:2023psd}, we can relate unbinned and binned correlators via the expectation of the idealized bispectrum estimator; this yields
\beq\label{eq: binning}
    [\mathcal{F}b]^{{\rm th},u}_{\chi}(\vec b) &\propto& \frac{1}{\Delta^u(\vec b)}\sum_{\ell_1\in b_1}\sum_{\ell_2\in b_2}\sum_{\ell_3\in b_3}\sqrt{\frac{(2\ell_1+1)(2\ell_2+1)(2\ell_3+1)}{4\pi}}\frac{1}{3}\left[\begin{pmatrix}\ell_1&\ell_2&\ell_3\\ -1&-1&2\end{pmatrix}+\text{2 perm.}\right]\\\nonumber
    &&\,\times\,\sum_{u'}S^{-1,u_1u_1'}_{\ell_1}S^{-1,u_2u_2'}_{\ell_2}S^{-1,u_3u_3'}_{\ell_3}\left.B_{\ell_1\ell_2\ell_3}^{u_1'u_2'u_3'}\right|_{\rm th}\\\nonumber
    \mathcal{F}^{{\rm th},u'u}_{\chi'\chi}(\vec b', \vec b)&\propto&\frac{1}{\Delta^u(\vec b)\Delta^{u'}(\vec b')}\sum_{\ell_1\in b_1}\sum_{\ell_2\in b_2}\sum_{\ell_3\in b_3}\frac{(2\ell_1+1)(2\ell_2+1)(2\ell_3+1)}{4\pi}\frac{1}{9}\left[\begin{pmatrix}\ell_1&\ell_2&\ell_3\\-1 & -1 &2\end{pmatrix}+\text{2 perms.}\right]^2\\\nonumber
    &&\,\times\,\left[S_{\ell_1}^{-1,u_1'u_1}S_{\ell_2}^{-1,u_2'u_2}S_{\ell_3}^{-1,u_3'u_3}\delta^{\rm K}_{b_1'b_1}\delta^{\rm K}_{b_2'b_2}\delta^{\rm K}_{b_3'b_3}+\text{5 perms.}\right]
\eeq
where we use weighting matrices $S_\ell^{XY}$ and assume that the theoretical bispectrum was specified by the isotropic form (which can be extracted from \eqref{eq: cmb-bispectra}, noting that the Dirac delta can be recast as a $3j$ symbol, after angular integration)
\beq
    \left.\av{a_{\ell_1m_1}^{X}a_{\ell_2m_2}^Ya_{\ell_3m_3}^Z}\right|_{\rm th}=\tj{\ell_1}{\ell_2}{\ell_3}{m_1}{m_2}{m_3}\left.B_{\ell_1\ell_2\ell_3}^{XYZ}\right|_{\rm th}.
\eeq
Here, the $u'$ summations in $[\mathcal{F}b]^{\rm th}$ extend over all 27 triplets of fields (not just those in \eqref{eq: field-triplets}, and $\Delta^u(\vec b)$ gives the degeneracy of the bin: $6$ if $u_1=u_2=u_3$ and $b_1=b_2=b_3$, $2$ if $u_1=u_2$ and $b_1=b_2$, $2$ if $u_2=u_3$ and $b_2=b_3$, or $1$ else. Finally, we implictly restrict to $(-1)^{\ell_1+\ell_2+\ell_3+n_B}=\chi$ in each case, and note that terms with $\chi\neq \chi'$ are independent, \textit{i.e.}\ bispectra sourced by parity-even and parity-odd physics are independent. Here, the complexity of the normalization appears due to the non-zero $TE$ cross-spectra; if this vanishes, $\mathcal{F}^{\rm th}$ is diagonal, and the action of the above is just to appropriately average the theory across each $\ell$-bin. Note that one could alternatively drop the $S^{-1}$ terms in both the estimator and the binned theory, computing an \textit{unweighted} average over all modes in the bin; from a Fisher forecast, this was found to increase the error-bar on $T$-mode parameter constraints by up to $45\%$ (or $5\%$ for $B$-modes), showing the utility of our weighting scheme.

\subsection{Covariances \& Likelihoods}\label{subsec: analysis-lik}
\noindent Finally, we consider how to constrain theoretical models (and non-Gaussianity amplitudes) from the data. As discussed in \citep{Philcox:2023ypl}, it is necessary to make a theoretical \textit{ansatz} for the covariance of the dataset, $\widehat{b}^u_\chi(\vec b)$, since the full covariance has $\simeq 6000^2$ elements, and cannot be robustly estimated from the $500$ simulations. As in previous works, we will assume that the correlation structure of $\mathrm{Cov}(\hat b,\hat b')$ is well described by the inverse normalization matrix, $\mathcal{F}^{-1}$; theoretically, this is appropriate since $\mathcal{F}^{-1}$ \textit{is} the covariance if the estimator is optimal, and empirically, it is found to be an excellent approximation, as demonstrated in \S\ref{sec: data}. In this approach, we first (reversibly) project the data and theory onto the Cholesky factorization of $\mathcal{F}$ \citep[e.g.,][]{Hamilton:1999uw}, defining
\beq\label{eq: beta-def}
    \beta^u_\chi(\vec b) \equiv \left[\mathcal{F}^{\rm T/2}b\right]^u_\chi(\vec b).
\eeq
and assume the following (diagonal) likelihood:
\beq\label{eq: likelihood}
    \mathcal{L}\big(\widehat{b}\big|b^{\rm th}\big) \propto \mathrm{exp}\left(-\frac{1}{2}\chi^2\big(\widehat{b}\big|b^{\rm th}\big)\right), \qquad \chi^2\big(\widehat{b}\big|b^{\rm th}\big)=\sum_{u,\chi,\vec b}\frac{\left[\widehat{\beta}^u_\chi(\vec b)-\beta^{{\rm th},u}_\chi(\vec b)\right]^2}{\mathrm{var}\left[\beta^u_\chi(\vec b)\right]}
\eeq
where $b^{\rm th}$ is some theoretical model, and the variance is estimated from the FFP10 realizations. If one works with only a subset of the data, it is important to take this subset before the rotation is applied (\textit{i.e.}\ to excise bins from $\mathcal{F}$ and $\widehat{b}$ rather than from $\widehat{\beta}$), else the results may incur bias. In the limit of an ideal estimator applied to Gaussian data, $\mathrm{var}\left[\beta^u_\chi(\vec b)\right]=1$; deviations from unity will thus encode non-Gaussianity from residual foregrounds and suboptimality in the treatment of the mask. 

The above likelihood allows for two tests of the data. Firstly, we can perform blind tests to assess whether there is \textit{any} evidence for a non-zero bispectrum of a given parity \citep[cf.,][]{PhilcoxCMB,Philcox:2023ypl,Philcox:2022hkh}. In this case, one sets $b^{\rm th}=0$ and compares the value of $\chi^2(\hat b|0)$ in the data (dropping bins of the wrong parity before projection) to the empirical distribution from simulations (or the theoretical expectation: a $\chi^2$ distribution with $N_{\rm bins}$ degrees-of-freedom). Secondly, we can constrain individual theoretical models, such as the gauge-field template given in \S\ref{subsec: theory-template}. Writing $b^{{\rm th},u}_\chi = f_{\rm NL}^{ttt}\,b^{{\rm template},u}_\chi$ for template $b^{\rm template}$ and amplitude $f_{\rm NL}^{ttt}$, constraints are obtained by sampling from the following posterior:
\beq\label{eq: posterior}
    \mathcal{P}\big(f_{\rm NL}^{ttt}\big|\hat{b}\big) \propto \mathrm{exp}\left(-\frac{1}{2}\chi^2\big(\hat b\big|f_{\rm NL}^{ttt}b^{\rm template}\big)\right).
\eeq
Often, one restricts to models of a given parity $\pm1$, which can be achieved by replacing $f_{\rm NL}^{ttt}\to f_{\rm NL}^{ttt,\pm1}$ and nulling bins in $b^{\rm template}$ with $\chi = \mp1$.\footnote{Since the rotation matrix can mix parities through the mask, this is not equivalent to restricting the summation of \eqref{eq: likelihood} to the relevant value of $\chi$.} We note that constraints obtained from a full theory model will generically be stronger than those from a blind test, since it allows for optimal model projection.

\section{Results}\label{sec: results}
\noindent We now present the main results of this paper: estimates of the \textit{Planck} $T$, $E$, and $B$-mode bispectra and constraints on the tensor-mode non-Gaussianity. We first examine the data itself in \S\ref{subsec: results-data}, before showing the model-independent blind-test constraints in \S\ref{subsec: results-blind}, and the constraints on $f_{\rm NL}^{ttt}$ in \S\ref{subsec: results-model}.

\begin{figure}
    \centering
    \includegraphics[width=0.95\textwidth]{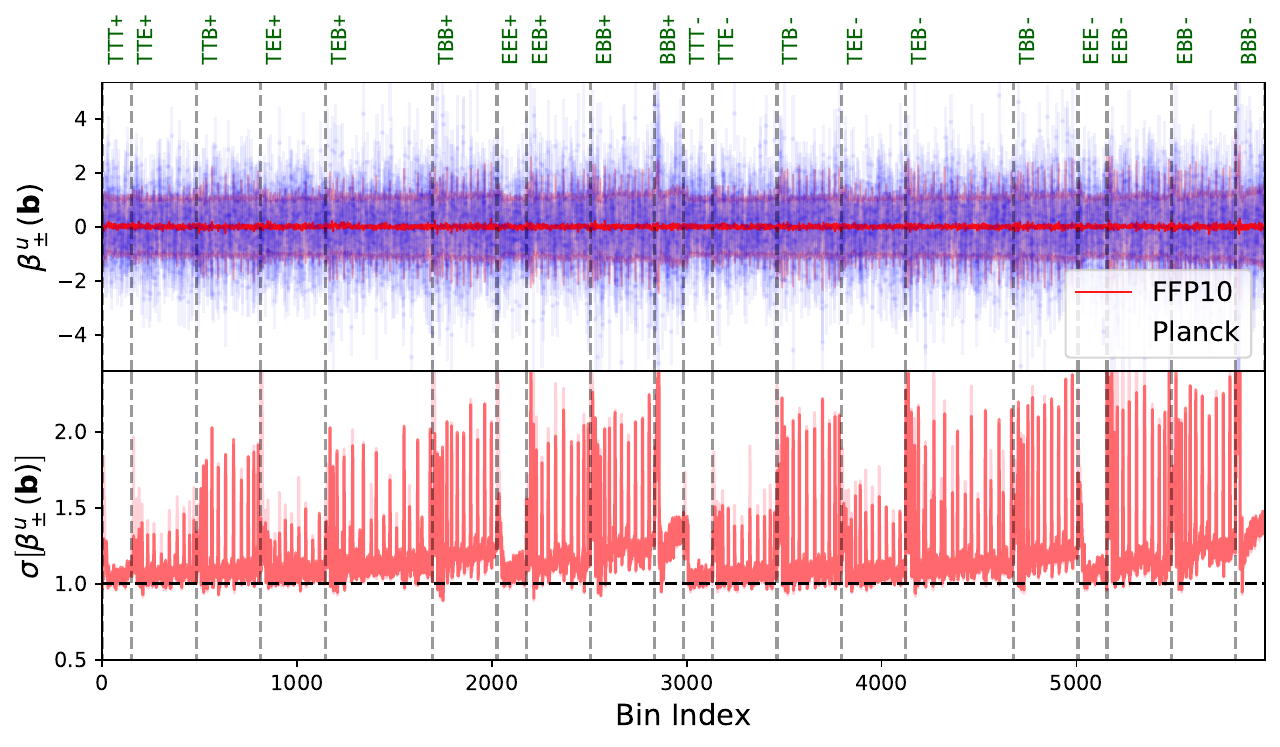}
    \caption{Large-scale binned bispectrum measurements from \textit{Planck} PR4 (blue points) compared to 500 FFP10 simulations (red bands) with $2\leq \ell< 500$, computed with the \textsc{PolyBin} code. We show the data (top) and error (bottom) for 20 combinations of bins and parities, separated by dashed lines and indicated in green. Spectra marked `+' are parity-symmetric, whilst those marked `-' are parity-antisymmetric (with `+' indicating even $\ell_1+\ell_2+\ell_3+n_B$, and `-' indicating the opposite, for $n_B$ $B$-modes in the statistic). In each subpanel, all bispectrum bins are stacked, with the smallest scales on the left-hand side, and we take the imaginary part of any spectra with odd $\ell_1+\ell_2+\ell_3$. We plot the rotated bispectrum defined in \eqref{eq: beta-def}; if the estimator is optimal and the data is Gaussian, this should be a unit normal variable. Here, we find slight enhancements in the variance (bottom panel), particularly for $\ell=2$ modes and polarization, indicating the effects of large-scale non-Gaussianity and foreground contamination. This will slightly enhance the variances of the parameter constraints, but will not lead to bias. Finally, the pink lines in the bottom panel show the effect of removing the linear term in the bispectrum estimator; we see a slight enhancement of the variance at low-$\ell$.}
    \label{fig: bispec-data}
\end{figure}

\subsection{Data}\label{subsec: results-data}
\noindent In Fig.\,\ref{fig: bispec-data}, we display the measured \textit{Planck} and FFP10 bispectra, projected as in \eqref{eq: beta-def}. Since this dataset contains almost $6000$ bins, visual assessment is difficult; however, the \textit{Planck} results do not appear to deviate wildly from the simulations, which themselves have averages highly consistent with zero (indicating that our pipeline does not contain significant additive bias). As mentioned above, if the data is Gaussian and the estimator is optimal, we would expect the variance of $\beta$ to be unity; here, we find some departure from this, which depends strongly on the triangle configuration (leading to the spikes in the figure). In general, we find slightly enhanced variances for triangles containing polarized low-$\ell$ modes; this indicates that our polarization estimators are somewhat sub-optimal, for example due to residual non-Gaussian foregrounds (as discussed below). Furthermore, we find that the large-scale variances are increased further if one drops the linear term in the bispectrum estimators (pink, cf.\,\ref{eq: rough-bispectrum-estimator}); this is as expected, and we consider its effects on the $f_{\rm NL}^{ttt}$ constraints in \S\ref{subsec: results-model}. We stress that $\sigma(\beta)>1$ does not imply that our estimator is biased, just that the variances will be slightly larger than would be obtained from an ideal analysis.

\begin{figure}
    \centering
    \includegraphics[width=0.48\textwidth]{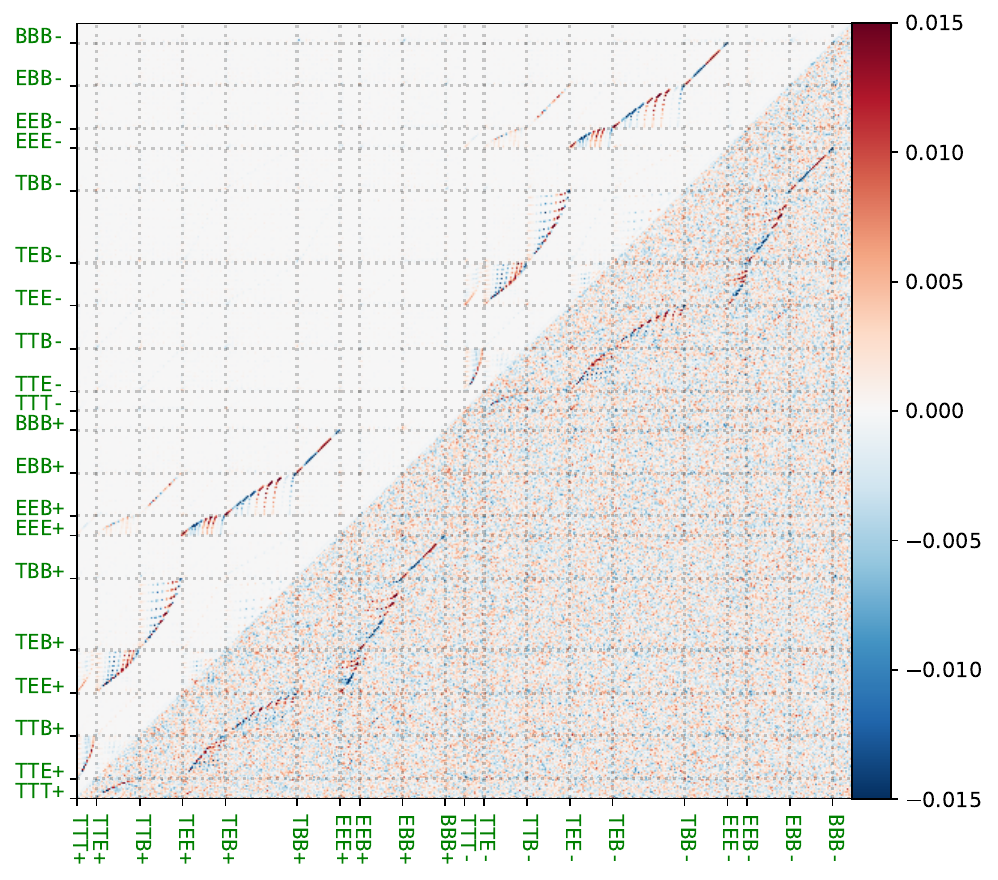}
    \includegraphics[width=0.48\textwidth]{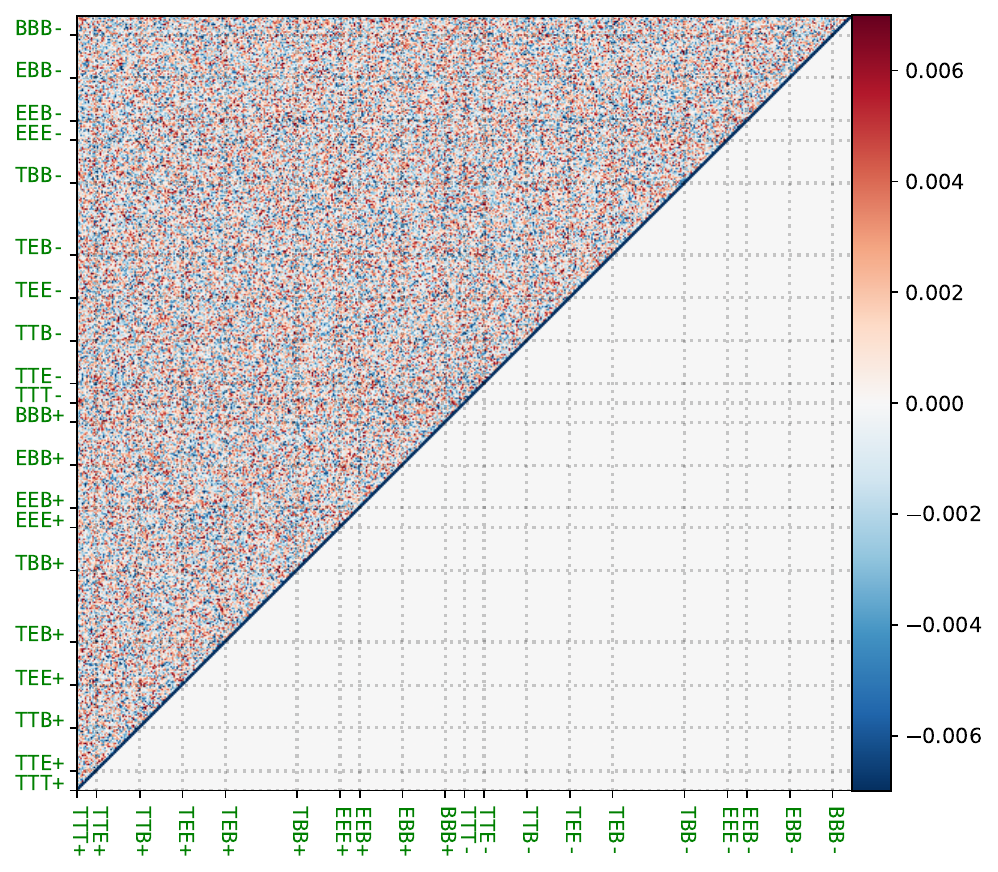}
    \caption{\textbf{Left:} Diagonal-subtracted correlation matrix of the binned bispectrum, $b^u_\chi(\vec b)$, obtained from theory (inverting $\mathcal{F}$, top left), and from 500 FFP10 simulations (bottom right), organizing components as in Fig.\,\ref{fig: bispec-data}. (Small) off-diagonal contributions arise from mask-induced leakage between different bins and parities and intrinsic correlations between $T$- and $E$-modes. \textbf{Right:} Correlation matrix for the projected $\beta^u_\chi(\vec b)$ statistic \eqref{eq: beta-def}, with the (symmetric) lower half removed). We find excellent agreement between the theory and mock $b^u_\chi(\vec b)$ covariance, indicating that the normalization matrix (which includes the mask and weighting scheme) captures the structure of the covariance well; this is evidenced by the covariance of $\tau$, which appears entirely diagonal.}
    \label{fig: bispec-corr}
\end{figure}

To more thoroughly assess the covariance of our dataset, we turn to the correlation matrix of the (unprojected) bispectrum, defined as
\beq
    \mathrm{Corr}\left(\widehat{b}_i,\widehat{b}_j\right) = \frac{\mathrm{Cov}\left(\widehat{b}_i,\widehat{b}_j\right)}{\sqrt{\mathrm{Var}\left(\widehat{b}_i\right)\mathrm{Var}\left(\widehat{b}_j\right)}},
\eeq
where $i,j$ jointly indexes the angular bin, field, and parity. This is shown in the left panel of Fig.\,\ref{fig: bispec-corr}, both for the theory (inverting the \textsc{PolyBin} normalization matrix $\mathcal{F}$) and for the FFP10 numerical covariance. In both cases, we find a roughly diagonal structure, but with non-trivial correlations at the $\simeq 2\%$ level between different fields, resulting from the non-zero $TE$ power spectrum. If one zooms in, smaller correlations (at the $\sim 0.5\%$ level) can be observed within the configuration of a given field (e.g. $b^{TTT}$); these arise from mask-induced leakage across $\ell$-bins. Although these effects are small (due to the broad bins used herein), they could lead to biases in our bispectrum estimates if uncorrected; in our case, the \textsc{PolyBin} normalization takes care of such complexities, returning an unbiased estimate of the binned spectrum.

In \S\ref{subsec: analysis-lik}, we claimed that the inverse normalization matrix, $\mathcal{F}^{-1}$, could be used to decorrelate the bins in the bispectrum data-vector, such that the projected statistic, $\beta$, would have diagonal covariance. From the left panel of Fig.\,\ref{fig: bispec-corr}, this assumption appears excellent, with the theoretical correlation structure well appearing highly consistent with that of the mocks. This is quantified in the right panel of Fig.\,\ref{fig: bispec-corr}, where we instead plot the correlation structure of $\widehat{\beta}$. As required, this is highly diagonal, with deviations fully consistent with noise,\footnote{The mean pixel value is $-5.3\times 10^{-6}$ with a standard deviation of $0.045$, against an expected value of $0.000\pm0.045$.} implying that our assumption of a diagonal likelihood \eqref{eq: likelihood} is robust, even in the presence of residual foregrounds found within FFP10.

\begin{figure}
    \centering
    \includegraphics[width=0.8\textwidth]{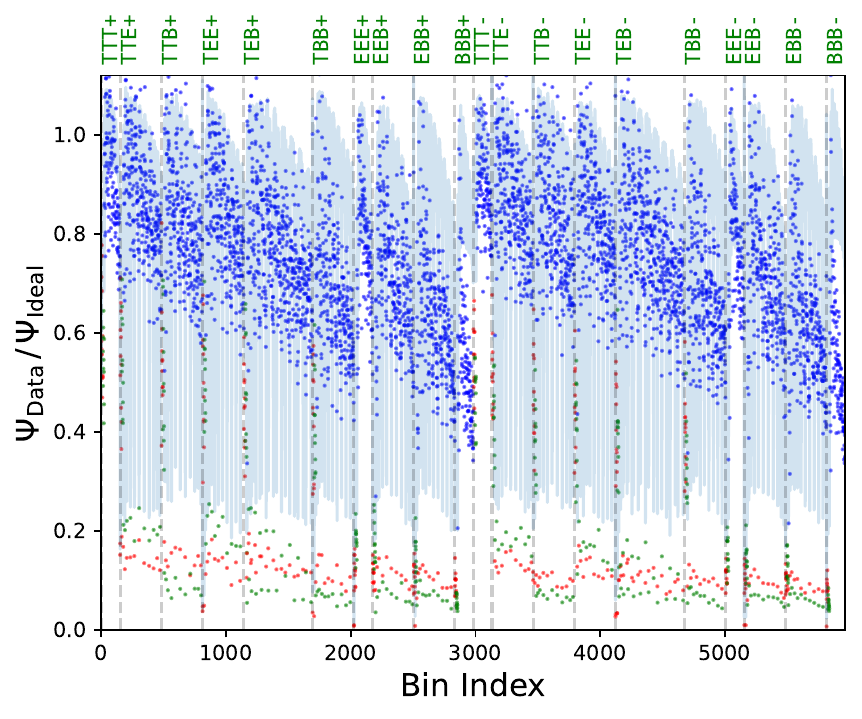}
    \caption{Ratio of true to ideal inverse bispectrum variances (denoted $\Psi \equiv \mathrm{diag}\left[\mathsf{C}^{-1}\right]$, assuming $f_{\rm sky}=0.78$). The points show the ratios obtained from the FFP10 inverse variances, whilst the lines show those predicted by the inverse normalization matrix, $\mathcal{F}^{-1}$. Red and green points indicate triangles containing modes with $\ell=2$ or $\ell=3$ respectively. Non-uniformities in the mask in the strongly reduce the precision of squeezed bispectrum measurements, as shown by the theory curves (which include the window); such triangles, and $B$-mode spectra, are further significantly impacted by non-Gaussian foreground residuals present in the FFP10 covariance.}
    \label{fig: variance-ratio}
\end{figure}

Finally, we turn to the (unprojected) bispectrum variances. In Fig.\,\ref{fig: bispec-theory} we plot the variance obtained from theory and simulations, finding excellent agreement between theory and simulations across many orders of magnitudes (as discussed above). Furthermore, we can assess the behavior of $\mathrm{var}(\widehat{b})$ for various bispectrum parameters: the behavior is broadly similar between the parity-even and -odd configurations, particularly at large $\ell$, though a few bins have (almost) vanishingly small covariance, since very few modes can populate them whilst obeying the restrictions on the sign of $\ell_1+\ell_2+\ell_3$. Different fields also have different variance structures: this occurs due to the very different signal-plus-noise power spectra, with, for example, the $B$-modes being dominated by almost scale-invariant noise. To assess the impact of masks and non-Gaussianity on this covariance, we show the ratio of the inverse covariance diagonal to that predicted from idealized theory (from the diagonal of the idealized $\mathcal{F}$ given in\,\ref{eq: binning}) in Fig.\,\ref{fig: variance-ratio}. The \textsc{PolyBin} estimate for the covariance encodes the distortions induced by the mask, but assumes Gaussianity; here, we find that such distortions have a scale-dependent effect on the inverse variances, which lead to considerable degradation in the constraints on triangles containing small $\ell$ (at a similar level for each correlator). The FFP10 results include also non-Gaussian contributions from noise and foregrounds; these are seen to significantly inflate the covariance of $B$-modes and large-scale triangles. Such effects are difficult to remove, but will be lessened in the future by higher experimental sensitivity and better foreground deprojection schemes.

\subsection{Blind Tests}\label{subsec: results-blind}
\noindent Armed with the \textit{Planck} dataset and the FFP10 simulations, we may ask a more generic question: does our dataset contain ``any'' evidence for bispectrum non-Gaussianity? To answer this, we perform blind tests similar to those used to constrain scalar parity-violation in the CMB and LSS four-point functions \citep{PhilcoxCMB,Philcox:2023ypl,Philcox:2022hkh}. This follows the methodology summarized in \S\ref{subsec: analysis-lik}; in brief, we compute a $\chi^2$ parameter from the \textit{Planck} data, relative to null assumptions, and compare its value to the empirical distribution extracted from simulations. For this purpose, we consider the parity-even and parity-odd sectors separately, and perform the test for a variety of data cuts, primarily focusing on $T+E+B$ and $T+E$ correlators (with the latter used in \citep{Planck:2019kim}). 

\begin{figure}
    \centering
    \includegraphics[width=0.7\textwidth]{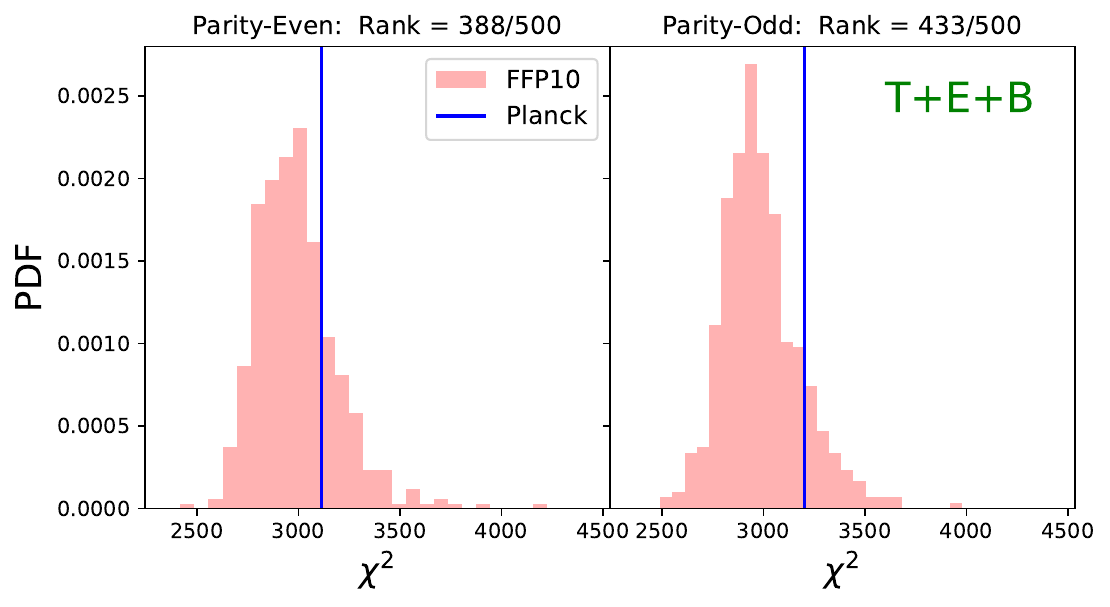}\\
    \includegraphics[width=0.7\textwidth]{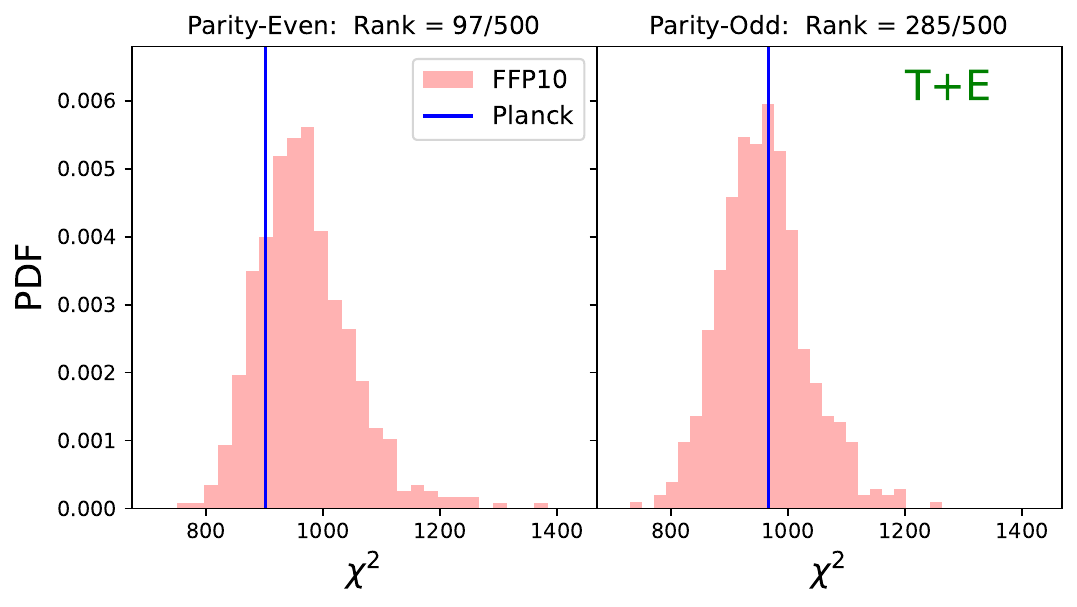}
    \caption{Blind tests for bispectra from the \textit{Planck} PR4 dataset. The left and right panels show the constraints on parity-even and parity-odd bispectra respectively, with the bottom panels displaying the results without $B$-modes. In all cases, we compute the $\chi^2$ statistic \eqref{eq: likelihood}, assuming zero theoretical model, and assign a rank to the \textit{Planck} dataset (blue) based on the distribution of values from 500 FFP10 simulations (red), \textit{i.e.}\ we perform a rank test. The \textit{Planck} data is consistent with noise within $1.1\sigma$; we find no evidence for non-Gaussianity in this model-independent test.}
    \label{fig: blind-test}
\end{figure}

\begin{table}[htb]
    \centering
    \begin{tabular}{l|cc}
    \textbf{Analysis} & Parity-Even & Parity-Odd\\\hline\hline
Fiducial ($T+E+B$) & $0.8\sigma$ & $1.1\sigma$\\\hline
$T$ only & $-0.4\sigma$ & $-1.2\sigma$\\
$E$ only & $-1.2\sigma$ & $0.4\sigma$\\
$B$ only & $1.7\sigma$ & $1.5\sigma$\\
$T+E$ & $-0.9\sigma$ & $0.2\sigma$\\
$T+B$ & $0.4\sigma$ & $1.0\sigma$\\
$E+B$ & $1.2\sigma$ & $1.3\sigma$\\\hline
$\ell_{\mathrm{min}}=4$ & $1.8\sigma$ & $2.3\sigma$\\
$\ell_{\mathrm{max}}=375$ & $0.8\sigma$ & $1.3\sigma$
\end{tabular}
    \caption{Model-independent constraints on tensor non-Gaussianity and parity-violation using \textit{Planck} PR4 bispectrum data with various analysis choices. In each case, we compare the \textit{Planck} $\chi^2$ value (specified in \eqref{eq: likelihood}) to the empirical distribution obtained from 500 FFP10 simulations, performing a rank test. Results are given in equivalent Gaussian $\sigma$, noting that only positive detections are physical. The fiducial analysis assumes $\ell_{\rm min}=2$ and $\ell_{\rm max}=500$. All detection significances are below $2.3\sigma$, indicating no compelling evidence for new physics.}
    \label{tab: blind-results}
\end{table}

The corresponding results are shown in Fig.\,\ref{fig: blind-test}\,\&\,Tab.\,\ref{tab: blind-results}. When $B$-modes are included in the analysis, the empirical $\chi^2$ distribution clearly departs from the (ideal) $\chi^2$-distribution; for this reason, we do not compare the \textit{Planck} data to a theoretical noise distribution. This likely arises from non-Gaussian residual foregrounds at low-$\ell$, and is significantly reduced by removing $B$-modes or increasing $\ell_{\rm max}$ to $4$. For the parity-even sector, the \textit{Planck} data is consistent with the simulations at $0.8\sigma$; the result is similar ($(-)0.9\sigma$) if $B$-modes are removed. This is an indirect constraint on both scalar and tensor non-Gaussianity, both of which can source parity-conserving physics. For parity-odd bispectra, we find consistency at $1.1\sigma$ ($0.2\sigma$) with (without) $B$-modes. This scenario is not sensitive to scalar physics; as argued in \citep{Coulton:2023oug} (and many other works), parity-odd bispectra can be sourced only by vector and tensor-mode physics. These conclusions are stable across an array of analysis choices (though the precise significances vary due to differing noise realizations): we detect no signal in any combination of $T$-, $E$- and $B$-modes and the results are broadly insensitive to scale cuts, with a maximum detection significance of $2.3\sigma$. As such, \textbf{we find no compelling evidence for tensor non-Gaussianity or parity-violation}.

Whilst we do not detect any signal in this study, we should make clear that this is \textit{not} an optimal test for specific models of non-Gaussianity; a better constraint can always be obtained in a model-specific analysis, as we perform below for the template given in \eqref{eq: hhh_fNLttt}. However, since the current work is the first to consider generic $B$-mode bispectra from observational data, it is pertinent to perform a general study in addition to model-specific constraints.

\subsection{Model Constraints}\label{subsec: results-model}

\begin{figure}
    \centering
    \includegraphics[width=0.85\textwidth]{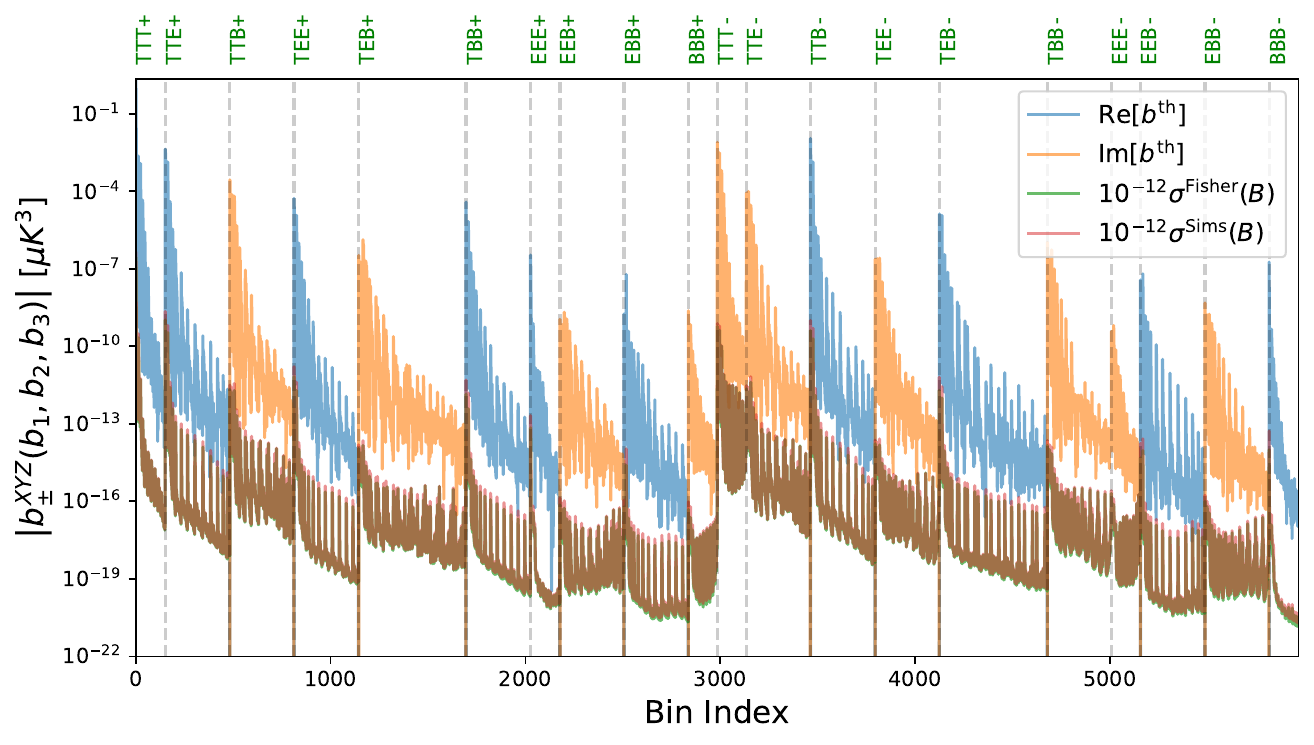}
    \caption{Comparison of theoretical tensor-mode non-Gaussianities (blue and orange) to the empirical and theoretical bispectrum variances (green and red), plotting as in Fig.\,\ref{fig: bispec-data}. The theoretical curves are computed with $f_{\rm NL}^{ttt,\pm}=1$ and we separate real (blue) and imaginary (orange) contributions. Here, modes generated by parity-even physics are shown on the left (`XYZ+' components), whilst those from parity-odd physics are on the right (`XYZ-'). We find excellent agreement between simulated (red) and predicted (green) variances across a wide range of scales, with the (projected) ratio plotted in Fig.\,\ref{fig: bispec-data}.}
    \label{fig: bispec-theory}
\end{figure}

\subsubsection{Theoretical Spectra}

\noindent Finally, we obtain constraints on the graviton bispectrum template specified in \eqref{eq: hhh_fNLttt}. Before constraining the amplitude parameter $f_{\rm NL}^{ttt}$, we briefly discuss the form of the binned spectra. As shown in Fig.\,\ref{fig: bispec-theory}, both parity-even and parity-odd components of the theory source real and imaginary parts; this is due to the presence of $B$-modes, and follows the logic described in \S\ref{subsec: bispectra}.
%; if one excludes $B$-modes from the analysis, parity-even physics generates real spectra with even $\ell_1+\ell_2+\ell_3$ (blue) whilst parity-odd physics creates imaginary spectra with odd $\ell_1+\ell_2+\ell_3$ (orange). With the curl-like polarization component, this logic is mixed up, thus we will henceforth refer only to the parity (the $\pm$ index in Fig.\,\ref{fig: bispec-theory}. 
We find strong scale-dependence of the bispectrum signal, with largest amplitudes observed for the lowest $\ell$-modes. There is some dependence on field, with $B$-modes having a lower amplitude; however, $B$-mode spectra do not suffer from cosmic-variance, thus their inclusion is expected to considerably sharpen constraints on $f_{\rm NL}^{ttt,\pm}$. Finally, we note similar forms for the parity-odd and parity-even physics; this is as expected, since the total number of modes with odd and even $\ell_1+\ell_2+\ell_3$ are similar. In contrast the variance of modes with even and odd $\ell_1+\ell_2+\ell_3$ can differ considerably (e.g., $EEE+$ has much lower variance than $EEE-$, with opposite behavior for $BBB+$ and $BBB-$); this is due to the Wigner $3j$ factors (which cancel in the theory model, but not in the bispectrum variance), who asymptotically vanish for $\ell_1+\ell_2+\ell_3$; \textit{i.e.}\ any signal is eventually washed out by projection effects.

\begin{figure}
    \centering
    \includegraphics[width=0.6\textwidth]{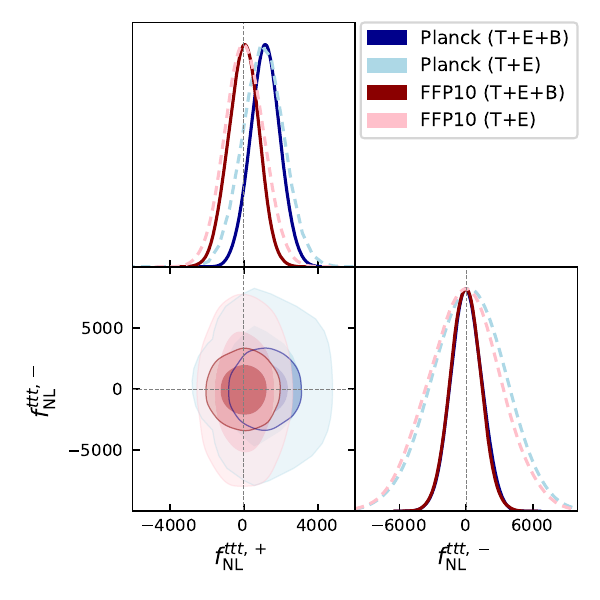}
    \caption{Constraints on parity-even and parity-odd gravitational wave bispectra from the \textit{Planck} PR4 temperature and polarization dataset (blue), as well as the mean of 500 FFP10 simulations (red). We show results using both $T+E$ (dashed lines, similar to \citep{Planck:2019kim}), and with the inclusion of $B$-modes (full lines), which is new to this work. $B$-modes improve the parity-even bounds by $\simeq 30\%$, and source the bulk of the parity-odd constraints. We find $f_{\rm NL}^{ttt,\pm}$ consistent with zero within $95\%$ confidence level in all cases. Numerical constraints are given in Tabs.\,\ref{tab: results}\,\&\,\ref{tab: results-ffp10}, and all results are obtained by sampling the posterior \eqref{eq: posterior} using \textsc{emcee} \citep{Foreman-Mackey:2012any}.}
    \label{fig: corner}
\end{figure}

\begin{table}[htb]
    \centering
    \begin{tabular}{l|ccc}
    \textbf{Analysis} & $\boldsymbol{10^{-2}f_{\rm NL}^{ttt,+}}$ & $\boldsymbol{10^{-2}f_{\rm NL}^{ttt,-}}$ & $\boldsymbol{10^{-2}f_{\rm NL}^{ttt}}$\\\hline\hline
    Fiducial ($T+E+B$) & $11 \pm 8$ & $-0 \pm 14$ & $9 \pm 7$\\\hline
$T$ only & $-2 \pm 20$ & $149 \pm 136$ & $5 \pm 20$\\
$E$ only & $67 \pm 30$ & $-384 \pm 439$ & $65 \pm 30$\\
$B$ only & $55 \pm 68$ & $-59 \pm 59$ & $-10 \pm 44$\\
$T+E$ & $10 \pm 11$ & $2 \pm 32$ & $10 \pm 10$\\
$T+B$ & $4 \pm 14$ & $-8 \pm 18$ & $-0 \pm 11$\\
$E+B$ & $40 \pm 17$ & $-20 \pm 40$ & $37 \pm 15$\\\hline
$\ell_{\mathrm{min}}=4$ & $13 \pm 9$ & $-6 \pm 15$ & $8 \pm 8$\\
$\ell_{\mathrm{max}}=375$ & $12 \pm 8$ & $0 \pm 14$ & $9 \pm 7$\\
No linear term & $12 \pm 8$ & $-1 \pm 14$ & $9 \pm 7$
    \\\hline
    $T$ only (\textit{Planck} 2018) & $4 \pm 17$ & $90 \pm 100$ & $6 \pm 16$\\\nonumber
    $E$ only (\textit{Planck} 2018) & $75 \pm 75$ & $-790 \pm 830$ & $70 \pm 75$\\\nonumber
    $T+E$ (\textit{Planck} 2018) & $16 \pm 14$ & $2 \pm 20$ & $13 \pm 12$
    \end{tabular}
    \caption{\textit{Planck} PR4 constraints on the equilateral tensor-mode non-Gaussianity parameter, $f_{\rm NL}^{ttt}$. For each analysis, we analyze parity-even (first column) and parity-odd (second column) bispectra separately, and additionally provide a combined constraint (third column), assuming maximal chirality. The fiducial analysis assumes $\ell_{\rm min}=2$ and $\ell_{\rm max}=500$, and we show the results from \textit{Planck} 2018 \citep{Planck:2019kim} in the bottom panel (assuming the \textsc{sevem} component separation pipeline). Most detection significances are small, with largest signals seen for $E$-modes or $E+B$-modes (up to $2.4\sigma$), though this is severely washed out when $T$-modes are included.}
    \label{tab: results}
\end{table}

\subsubsection{Main Constraints}
\noindent In Fig.\,\ref{fig: corner}\,\&\,Tab.\,\ref{tab: results}, we present constraints on the tensor non-Gaussianity amplitudes $f_{\rm NL}^{ttt,\pm}$ from \textit{Planck} $T$, $E$ and (optionally) $B$-mode data. In our fiducial analysis, including all polarization types, we find $f_{\rm NL}^{ttt,+}=(11\pm8)\times 10^2$ and $f_{\rm NL}^{ttt,-}=(0\pm14)\times 10^2$, both of which are consistent with zero at $1.4\sigma$. As such, \textbf{we report no detection of equilateral tensor non-Gaussianity, both in the parity-even and parity-odd sector}. The two constraints are almost entirely uncorrelated; this is as expected, since the parity-odd and parity-even sectors do not mix in an ideal bispectrum estimator \citep{Philcox:2023psd}, and the mask-induced leakage between different parities is seen to be small in Fig.\,\ref{fig: bispec-corr}. Assuming $f_{\rm NL}^{ttt,+}=f_{\rm NL}^{ttt,-}$ (true for a maximally chiral gravitational wave sector), we obtain the combined constraint $f_{\rm NL}^{ttt}=(9\pm 7)\times 10^2$, which is dominated by the parity-even sector.

Excluding $B$-modes from the analysis is found to inflate the parity-even constraints by $\simeq 30\%$, whilst increasing the error-bar on $f_{\rm NL}^{ttt,-}$ by a factor $\simeq 2.4$. In the absence of $B$-modes, parity-odd physics can be probed only through modes with odd $\ell_1+\ell_2+\ell_3$, which, by intermediate $\ell$, are limited by projection effects. In contrast, $TTB$, $TEB$, $EEB$ and $BBB$ spectra probe parity-violation with even $\ell_1+\ell_2+\ell_3$, avoiding such a suppression, and sharpening the constraints. If we consider only $T$-modes, the constraints degrade significantly; this implies that much of the signal-to-noise comes from polarization and its cross-correlation with temperature (noting that the full dataset contains $10$ spectra). A similar story holds for analyses involving just $E$- or $B$-modes, with the lack of cross-spectra significantly reducing the signal-to-noise. For $E$-only or $E+B$-modes, we find that $f_{\rm NL}^{ttt,+}$ is non-zero at up to $2.4\sigma$ (possibly indicating residual systematic contamination): however, this is strongly inconsistent with the full dataset result. 

\begin{figure}
    \centering
    \includegraphics[width=0.49\textwidth]{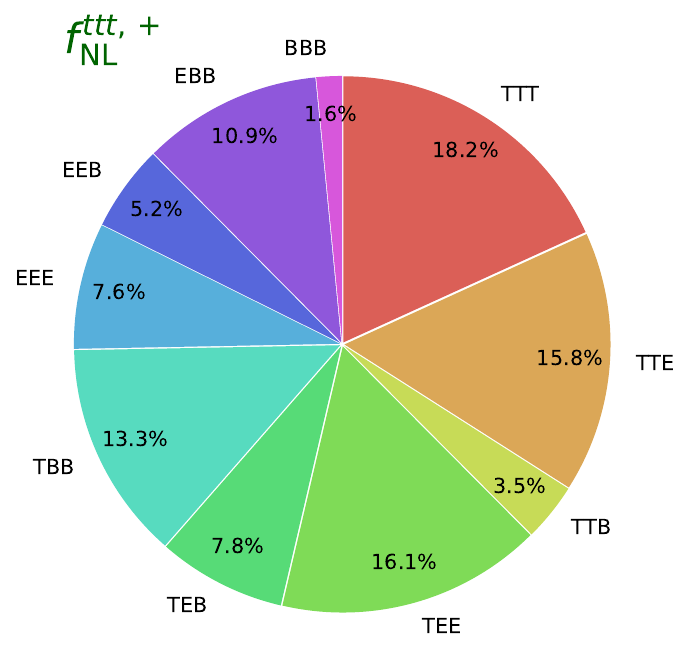}
    \includegraphics[width=0.49\textwidth]{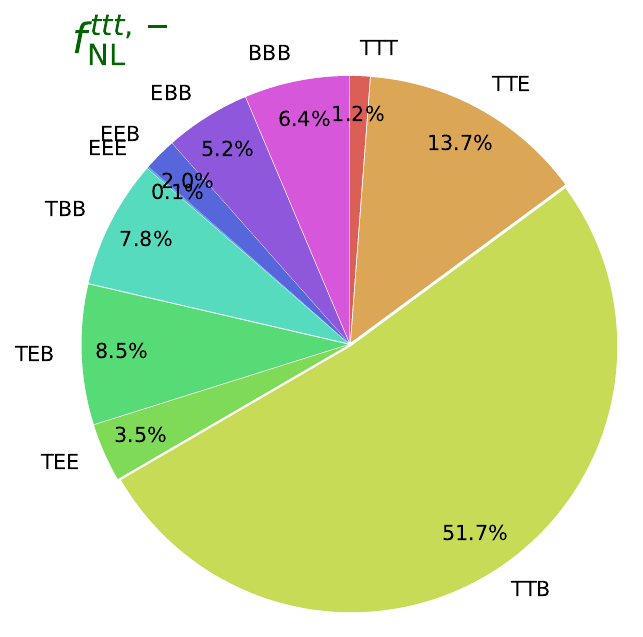}
    \caption{Fractional contribution of each bispectrum to the $f_{\rm NL}^{ttt,\pm}$ constraints, here parametrized by the Fisher information $\sigma_{XYZ}^{-2}(f_{\rm NL}^{ttt,\pm})$. We show results for the parity-even and parity-odd sector in the left and right panel, and perform all analyses using fiducial hyperparameters, with $\ell\in[2,500)$. We find that $T$- and $E$-modes dominate the parity-even sector (with non-negligible $TBB$ contributions), but the parity-odd sector is primarily influenced by $TTB$ (since it sources spectra with even $\ell_1+\ell_2+\ell_3$).}
    \label{fig: pie-plots}
\end{figure}

To robustly assess the contributions of each field to the overall constraint, we perform a likelihood analysis using each $b^{XYZ}$ bispectrum in turn, and plot the fractional contributions to $\sigma^{-2}(f_{\rm NL}^{ttt,\pm})$ (\textit{i.e.}\ the Fisher information) in Fig.\,\ref{fig: pie-plots}. For the parity-even sector, we find that constraints are split across a wide variety of channels, $T+E$-mode correlators make up $\simeq 60\%$ of the Fisher matrix, with the remainder sourced by $B$-mode spectra (particularly those with two $B$-modes, following the above projection arguments). This is quite different to the ratios found for equilateral \textit{scalar} non-Gaussianity in \citep{Planck:2019kim}, reflecting the utility of spin-two $E$-modes in graviton constraints. Parity-odd constraints are instead strongly dominated by the $TTB$ correlator. This matches our expectation, since it is the lowest-noise correlator with even $\ell_1+\ell_2+\ell_3$; in this case, the utility of $B$-modes is abundantly clear, with $T$-only constraints making up only $1\%$ of the Fisher information. 

Finally, it is interesting to ask to what extent these constraints could be obtained from the model-independent analysis of \S\ref{subsec: results-blind}. In this case once can ask: how large would $f_{\rm NL}^{ttt,\pm}$ need to be to induce a $1\sigma$ deviation in the \textit{Planck} $\chi^2$ value relative to the empirical distribution from FFP10? This is straightforward to compute and leads to the error-bars $\sigma(f_{\rm NL}^{ttt,+}) = 1.3\times 10^4$, $\sigma(f_{\rm NL}^{ttt,-})=1.9\times 10^4$ from the fiducial analysis including $T$, $E$, and $B$-modes. These constraints are \textit{much} weaker than those given in Tab.\,\ref{tab: results}, re-emphasizing our earlier statement that model-specific tests always obtain tighter parameter constraints than blind challenges.

\subsubsection{Consistency Checks}

\begin{figure}
    \centering
    \includegraphics[width=0.95\textwidth]{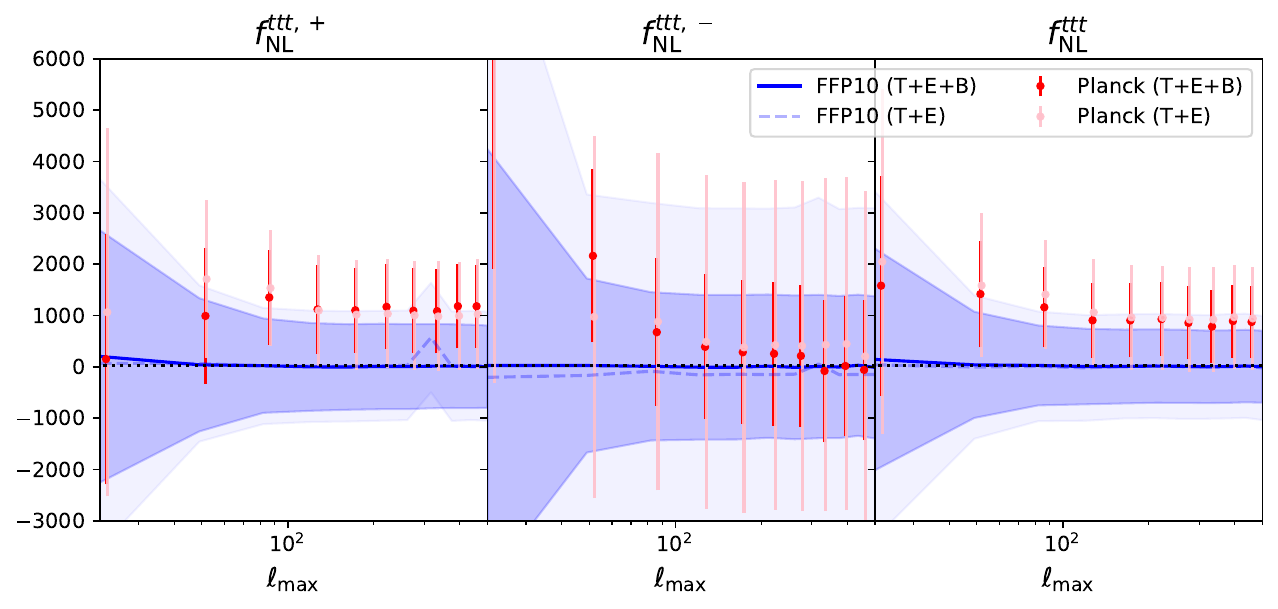}
    \caption{Constraints on the parity-even (left), parity-odd (center) and combined (right) tensor non-Gaussianity parameters as a function of the maximum scale used in the analysis, $\ell_{\rm max}$. Results from \textit{Planck} PR4 are shown as red data-points (or pink, dropping $B$-modes), whilst the blue bands indicate results obtained from analyzing the mean of 500 FFP10 simulations. As expected, the constraining power saturates by $\ell_{\rm max}\simeq 200$.}
    \label{fig: results-lmax}
\end{figure}

\noindent Next, we consider the dependence of our constraints on various hyperparameters (with results given in Tab.\,\ref{tab: results}). Firstly, we consider excising modes with $\ell\leq 4$; this inflates the error-bars by $\simeq 10\%$ but does not significantly shift the mean, implying that any bias induced by large-scale residual foregrounds is small. Secondly, we explore the dependence of the constraints on $\ell_{\rm max}$ both in Fig.\,\ref{fig: results-lmax}. Constraints from both the $T+E$ and $T+E+B$ datasets degrade significantly if we restrict only to very low-$\ell$; this occurs due to the form of the tensor transfer functions. Beyond $\ell_{\rm max}\simeq 200$, the constraints appear to stabilize; indeed, we find negligible changes when the analysis is repeated without the highest-two $\ell$-bins (Tab.\,\ref{tab: results}). This matches the forecasts of \citep{Shiraishi:2013kxa,Shiraishi:2019yux}, and is expected to be similar even for future low-noise experiments, due to the lensing $B$-mode floor. Different scalings are found for futuristic experiments when invoking delensing however, as well as for mixed scalar-scalar-tensor \citep{Shiraishi:2013vha,Domenech:2017kno,Shiraishi:2019yux,Duivenvoorden:2019ses} and scalar-tensor-tensor \citep{Shiraishi:2013vha,Bartolo:2018elp} bispectra. These scalings motivate our choice of $\ell_{\rm max}$ (and $N_{\rm side}$) used throughout this work, and indicates that one could perform faster analyses (or more optimally binned) with lower $\ell_{\rm max}$ in the future. Furthermore, the bispectrum estimation could be optimized by a factor $\simeq 100$: as seen in Tab.\,\ref{tab: results}, the linear term in the bispectrum estimator makes essentially no difference to the $f_{\rm NL}^{ttt,\pm}$ constraints and could be dropped.\footnote{This point is not generic; theoretical models whose signatures are concentrated at the lowest $\ell$-modes (such as $f^{\rm loc}_{\rm NL}$) would strongly benefit from this term.}

\begin{table}[htb]
    \centering
    \begin{tabular}{l|ccc}
    \textbf{Analysis} & $\boldsymbol{f_{\rm NL}^{ttt,+}}$ & $\boldsymbol{f_{\rm NL}^{ttt,-}}$ & $\boldsymbol{f_{\rm NL}^{ttt}}$\\\hline\hline
    Fiducial ($T+E+B$) & $0 \pm 8$ & $-0 \pm 14$ & $0 \pm 7$\\\hline
$T$ only & $1 \pm 20$ & $-1 \pm 135$ & $1 \pm 20$\\
$E$ only & $0 \pm 30$ & $-20 \pm 436$ & $0 \pm 30$\\
$B$ only & $-0 \pm 68$ & $-6 \pm 60$ & $-3 \pm 44$\\
$T+E$ & $-0 \pm 11$ & $-2 \pm 32$ & $-0 \pm 10$\\
$T+B$ & $0 \pm 14$ & $-0 \pm 18$ & $0 \pm 11$\\
$E+B$ & $-0 \pm 17$ & $-2 \pm 40$ & $-1 \pm 16$\\\hline
$\ell_{\mathrm{min}}=4$ & $0 \pm 9$ & $-0 \pm 15$ & $0 \pm 8$\\
$\ell_{\mathrm{max}}=375$ & $0 \pm 8$ & $-0 \pm 14$ & $0 \pm 7$\\
No linear term & $-1 \pm 8$ & $-0 \pm 14$ & $0 \pm 7$
        \end{tabular}
    \caption{As Tab.\,\ref{tab: results}, but for the mean of 500 FFP10 mocks. Results are consistent with zero, indicating that our pipeline is unbiased.}
    \label{tab: results-ffp10}
\end{table}

Lastly, we have performed extensive verification of our pipeline using the FFP10 simulation suite, leading to the constraints shown in Fig.\,\ref{fig: corner}\,\&\,Tab.\,\ref{tab: results-ffp10}. In every analysis, analyzing the mean of 500 simulations returns a result consistent with zero to within $0.2\sigma$. This indicates that the systematic error of our pipeline is under control, at least for the systematics included in the FFP10 simulation suite, and gives us confidence in the robustness of the above results. We further note that non-Gaussian contributions to the CMB bispectrum, namely ISW-lensing and polarization-lensing contributions, peak in squeezed regimes and at larger-$\ell$, and are thus not expected to bias our (large-scale and equilateral) constraints \citep{Hu:2000ee,Lewis:2011fk,Pearson:2012ba}.

\subsubsection{Interpretation}
\noindent The $T+E$ constraints on $f_{\rm NL}^{ttt,\pm}$ can be directly compared to those of the \textit{Planck} 2018 non-Gaussianity analysis \citep{Planck:2019kim} (shown in Tab.\,\ref{tab: results}); the \textsc{sevem} pipeline (which is also used in \textsc{npipe}) found $f_{\rm NL}^{ttt,+}=(16\pm 14)\times 10^2$ with $f_{\rm NL}^{ttt,-}=(2\pm 20)\times 10^2$; our parity-even constraint is $\approx 30\%$ tighter, though our parity-odd error-bar is degraded by a factor of $1.6$. To rationalize this, we note a number of key differences in the analysis. Firstly, our work utilizes an updated \textit{Planck} dataset with better treatment of foregrounds and low-$\ell$ polarization data. Secondly, we use lower $\ell_{\rm min}$ ($2$ rather than $4$, which gives a $\simeq 10\%$ improvement). Finally, our analysis utilizes binned bispectra in contrast to the modal approach \citep{Shiraishi:2014roa,Shiraishi:2019exr} used in \citep{Planck:2019kim}. Our $\ell$-bins are relatively broad (using only $13$ $\ell$-bins across a wide range of scales), which will lead to some degradation in constraining power. This is not necessarily a limitation of the approach however; if our analysis was repeated at lower $\ell_{\rm max}$ (justified by Fig.\,\ref{fig: results-lmax}) and without the costly linear term (justified by Tab.\,\ref{tab: results}), one could easily use thinner $\ell$-bins and thus capture this lost signal-to-noise; Fisher forecasts using $23$ $\ell$-bins indicate that the constraint on $f_{\rm NL}^{ttt,+}$ ($f_{\rm NL}^{ttt,-}$) could be improved by $10-15\%$. We note, however, that by the inclusion of $B$-modes, our constraints already represent a significant enhancement over those presented previously, thus this discussion simply indicates further improvements are possible.

\begin{figure}
    \centering
    \includegraphics[width=0.9\textwidth]{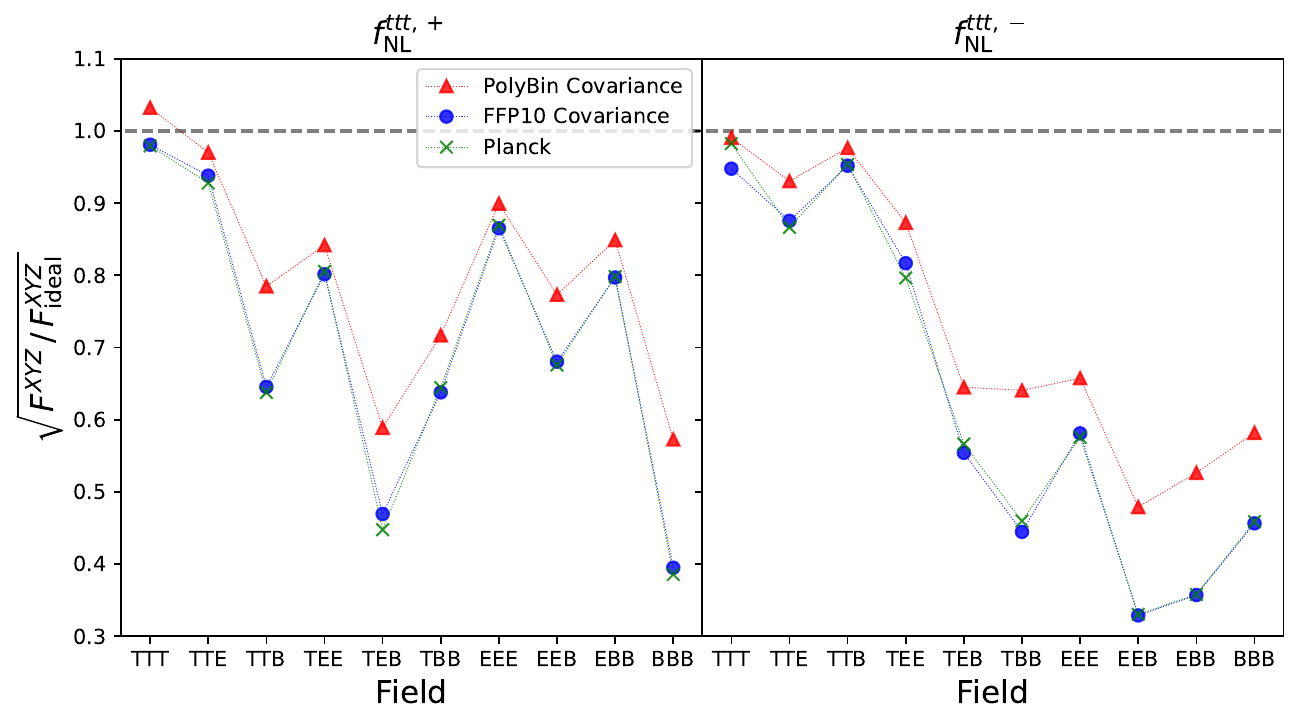}
    \caption{Ratio of the expected to idealized (inverse) error-bars on the non-Gaussianity parameter $f_{\rm NL}^{ttt,\pm}$ computed from the Fisher matrix of \eqref{eq: fish} with differing choices of covariance matrix. We show results with the expected covariance including the mask (red), that extracted from 500 FFP10 simulations (blue), and the true error-bars obtained from the likelihood analysis of \textit{Planck} (green). We conclude the the non-uniform window function degrades the constraints by $\sim 30\%$, particularly for polarization and parity-odd models, and there is a further $\sim 10\%$  reduction from non-Gaussian contributions to the polarization covariance, encoded within FFP10.}
    \label{fig: fish-ratio}
\end{figure}

Finally, it is interesting to compare our constraints to those expected from simplified Fisher forecasts. This is shown in Fig.\,\ref{fig: fish-ratio}, where we plot the single-field Fisher information:
\beq\label{eq: fish}
    F^{XYZ} = \sum_{\vec b,\vec b'}\frac{\partial b^{XYZ}(\vec b)}{\partial f_{\rm NL}^{ttt,\pm}}\left[\mathsf{C}^{XYZ}\right]^{-1}(\vec b,\vec b')\frac{\partial b^{XYZ}(\vec b')}{\partial f_{\rm NL}^{ttt,\pm}}
\eeq
for $X,Y,Z\in\{T,E,B\}$ and bins $\vec b,\vec b'$, relative to that expected from an ideal binned analysis with $f_{\rm sky}=0.78$ (using the covariance of \eqref{eq: binning}, as shown in Fig.\,\ref{fig: variance-ratio}). We consider three Fisher forecasts: (a) using the covariance extracted from \textsc{PolyBin} (inverting the normalization, $\mathcal{F}$), which fully encapsulates geometric distortions; (b) using the covariance extracted from FFP10 simulations, which further accounts for non-Gaussian contributions to the covariance; (c) replacing $F^{XYZ}$ with the error-bars extracted from the actual analysis of \textit{Planck} data. For $TTT$-bispectra, all ratios are close to unity, indicating that our methodology is optimal (given our binning) and that \textit{Planck} temperature is fully cosmic-variance limited. When polarization is included, the \textsc{PolyBin} constraints degrade by $\simeq 10-40\%$, particularly with the inclusion of $B$-modes or parity-odd sector. This indicates that the non-uniform mask significantly complicates our analysis due to its strong scale-dependence at low-$\ell$ (given that the expected $B$-mode signal is small). The FFP10 result (which agrees with \textit{Planck} bounds, as expected for a Gaussian posterior) shows that there is some impact of non-Gaussian covariance (\textit{i.e.}\ contributions to the covariance that are not sourced by the two-point function); this reduces the constraining power by $\simeq 10\%$. These results are in accordance with our discussion of the covariances themselves in \S\ref{subsec: results-data}. 

\section{Discussion}\label{sec: summary}
\noindent Primordial gravitational waves represent a unique probe of the early Universe, shedding light both on the inflationary vacuum and interactions within multi-field paradigms. There exist many sources of tensor fluctuations; to distinguish between them, one may consider their statistical properties, in particular their non-Gaussianity and parity. In this work, we have used the full \textit{Planck} temperature and polarization data to place novel constraints on graviton bispectra, considering both parity-even and parity-odd scenarios. Such correlators can be sourced by a number of mechanisms, including non-standard inflationary vacua, gravitational interactions of gauge fields, scalar-tensor theories of general relativity, non-attractor phases of inflation, Chern-Simons modified gravity, and beyond, and could, in principle be large compared to the two-point function %\citep{Cook:2013xea,Namba:2015gja,Dimastrogiovanni:2016fuu,Gong:2023kpe,Kanno:2022mkx,Ozsoy:2019slf,Mylova:2019jrj,DeLuca:2019jzc,Cabass:2021fnw,Cabass:2022jda,Bordin:2020eui,Cabass:2021iii,Pajer:2020wxk,Shiraishi:2011st,Bartolo:2017szm,Bartolo:2018elp}.
\citep[e.g.,][]{Namba:2015gja,Dimastrogiovanni:2016fuu,Gong:2023kpe,Ozsoy:2019slf,Mylova:2019jrj,Cabass:2021fnw,Cabass:2022jda,Bordin:2020eui,Cabass:2021iii,Shiraishi:2011st,Bartolo:2018elp}.
%[MS: Here, from the context, we should limit the citations to the works discussing visible parity-violating GW non-Gaussianity. I reduced the references.]}

Utilizing recently developed binned bispectrum estimators \citep{Philcox:2023uwe,Philcox:2023psd,PolyBin}, we have measured the binned \textit{Planck} $T$-, $E$- and $B$-mode bispectra across a range of scales, and used these to perform model-agnostic tests of graviton parity-violation and non-Gaussianity. Furthermore, we have placed constraints on a popular model of equilateral tensor non-Gaussianity (encoding the gravitational interactions of gauge fields), considering both the parity-even and parity-odd sectors. This work represents the first use of $B$-modes to probe non-Gaussian inflationary physics; we find that their inclusion enhances constraints on the parity-even sector by $\simeq 30\%$ and dominates parity-odd constraints, particularly through $TTB$ bispectra. Throughout our tests, we found no significant detections; as such, \textbf{we report no compelling evidence for gravitational wave bispectra, in both the parity-even and parity-odd sectors}.

Comparing to the WMAP, \textit{Planck} 2015 and \textit{Planck} 2018 analysis \citep{Shiraishi:2014ila,Planck:2015zfm,Planck:2019kim}, we find significantly enhanced constraints on the tensor non-Gaussianity parameter $f_{\rm NL}^{ttt}$, primarily due to our inclusion of $B$-modes. Without $B$-modes, our parity-odd constraints are somewhat worse (though our parity-even constraints improve over the former, due to lower $\ell_{\rm min}$). This is attributed to the wide $\ell$-bins used in our bispectrum estimator, and could be mollified by reducing the bin width; we forecast that constraints could be improved by $10-15\%$ by doubling the number of linear bins. As noted above, our estimator could be significantly expedited by dropping the linear term (which increases computation by $\simeq 100\times$ and is of little use for equilateral templates), and by reducing the maximum scale, noting that our constraints saturate by $\ell_{\rm max}\simeq 200$. We have also compared our results to idealized Fisher forecasts; here, we find that the scale-dependence of the mask considerably reduces the precision of $B$-mode analyses, and there is some non-Gaussian contribution to the covariance, which is well captured by the FFP10 simulations. Such effects will be reduced in the future by better foreground subtraction and understanding of the galactic mask.

Our constraints on tensor non-Gaussianity will improve considerably in the future. Whilst the large-scale $T$- and $E$-modes are quickly becoming cosmic-variance-limited, our work has shown that $B$-modes can significantly enhance both parity-even and parity-odd constraints. Future experiments such as the Simons Observatory, CMB-S4 and LiteBIRD, will see such modes measured with much higher precision, and improvements in delensing techniques are poised to reduce their remaining cosmic variance, allowing information to be extracted also from smaller scales. For example, the LiteBIRD survey is expected to reach an $f_{\rm NL}^{ttt}$ errorbar of $\mathcal{O}(1)$ from (lensed) $B$-modes alone (representing a $70\times$ improvement over the current constraint) \citep{Shiraishi:2019yux}. \resub{Furthermore, the techniques considered herein can be simply extended to other scenarios, such as local-type tensor non-Gaussianity or mixed tensor-scalar-scalar non-Gaussianity, which will result in tighter bounds on the corresponding amplitudes.} By including the full set of $T+E+B$-mode correlators, \resub{we can strengthen the bounds on a wide variety of primordial tensor phenomena}, which will be of paramount importance in future constraints on inflation.
\acknowledgments
{\small
\begingroup
\hypersetup{hidelinks}
\noindent 
\resub{We thank the anonymous referee for insightful comments.} OHEP is a Junior Fellow of the Simons Society of Fellows and thanks \href{https://www.reddit.com/r/funny/comments/1xht1z/a_catholic_priest_and_a_hilarious_llama/}{Lleviticus the Llama} for ecclesiastical advice. MS is supported by JSPS KAKENHI Grant Nos. JP20H05859 and JP23K03390. MS also acknowledges the Center for Computational Astrophysics, National Astronomical Observatory of Japan, for providing the computing resources of the Cray XC50.
\endgroup
}

\appendix

\bibliographystyle{apsrev4-1}
\bibliography{refs}% Produces the bibliography via BibTeX.

\end{document}